\documentclass{elsarticle}
\usepackage{amsmath}
\usepackage{amssymb}
\usepackage{units}
\usepackage{xspace}
\usepackage{subfig}
\usepackage{graphicx}
\usepackage{units}
\usepackage{url}
\usepackage{color}
\usepackage{ulem}

\def\EeV{\ifmmode {\mathrm{\ Ee\kern -0.1em V}}\else
                   \textrm{Ee\kern -0.1em V}\fi}%
\def\eV{\ifmmode {\mathrm{\ e\kern -0.1em V}}\else
                   \textrm{e\kern -0.1em V}\fi}%
\def\gcm{\ifmmode {\mathrm{g/cm}^2}\else
                   {g/cm$^2$}\fi\xspace}%
\def\Xmax{\ifmmode {X_\mathrm{max}}\else
                   {$X_\mathrm{max}$}\fi\xspace}%
\newcommand{\energy}[1]{\unit[$10^{#1}$]{\eV}}
\newcommand{\depth}[1]{\unit[#1]{\gcm}}

\begin{document}

\begin{frontmatter}
\title{The exposure of the hybrid detector of the Pierre Auger Observatory}

% This is a fragment that can be inserted into a LaTeX document:
\par\noindent
{\bf The Pierre Auger Collaboration} \\
P.~Abreu$^{73}$, 
M.~Aglietta$^{55}$, 
E.J.~Ahn$^{89}$, 
D.~Allard$^{31}$, 
I.~Allekotte$^{1}$, 
J.~Allen$^{92}$, 
J.~Alvarez Castillo$^{66}$, 
J.~Alvarez-Mu\~{n}iz$^{80}$, 
M.~Ambrosio$^{48}$, 
A.~Aminaei$^{67}$
% ...

\par\noindent
$^{1}$ Centro At\'{o}mico Bariloche and Instituto Balseiro (CNEA-
UNCuyo-CONICET), San Carlos de Bariloche, Argentina \\
% ...
$^{31}$ Laboratoire AstroParticule et Cosmologie (APC), 
Universit\'{e} Paris 7, CNRS-IN2P3, Paris, France \\
% ...
$^{48}$ Universit\`{a} di Napoli "Federico II" and Sezione INFN, 
Napoli, Italy \\
% ...
$^{55}$ Istituto di Fisica dello Spazio Interplanetario (INAF),
 Universit\`{a} di Torino and Sezione INFN, Torino, Italy \\
% ...
$^{66}$ Universidad Nacional Autonoma de Mexico, Mexico, D.F., 
Mexico \\
$^{67}$ IMAPP, Radboud University, Nijmegen, Netherlands \\
% ...
$^{73}$ LIP and Instituto Superior T\'{e}cnico, Lisboa, Portugal \\
% ...
$^{80}$ Universidad de Santiago de Compostela, Spain \\
% ...
$^{89}$ Fermilab, Batavia, IL, USA \\
% ...
$^{92}$ New York University, New York, NY, USA \\
% ...
\par\noindent
($\dagger$) Deceased \\
(a) at Konan University, Kobe, Japan \\
(c) at Caltech, Pasadena, USA \\
% last updated:	6/25/2010 

\begin{abstract}
The Pierre Auger Observatory is a detector for ultra-high energy
cosmic rays. It consists of a surface array to measure secondary
particles at ground level and a fluorescence detector to measure the
development of air showers in the atmosphere above the array.  The
``hybrid'' detection mode combines the information from the two
subsystems. We describe the determination of the hybrid exposure for
events observed by the fluorescence telescopes in coincidence with at
least one water-Cherenkov detector of the surface array. A detailed
knowledge of the time dependence of the detection operations is
crucial for an accurate evaluation of the exposure.  We discuss the
relevance of monitoring data collected during operations, such as the
status of the fluorescence detector, background light and atmospheric
conditions, that are used in both simulation and reconstruction.
\end{abstract}

\end{frontmatter}

\section{Introduction}
\label{sec:intro}

The Pierre Auger Observatory has been designed to investigate the
origin and the nature of Ultra High Energy Cosmic Rays.  It consists
of a large array of about 1600 surface stations (the {\it SD array})
covering an area of \unit[3000]{km$^2$} for detecting the secondary
particles of the air shower at ground level by means of the Cherenkov
radiation they produce in water.  The ground array is overlooked by 24
air fluorescence telescopes (the {\it FD system}), grouped in 4
enclosures each consisting of 6 optical telescopes. These devices are
used to observe the longitudinal profile of cosmic ray showers on
clear moonless nights.  The Observatory, located outside the town of
Malarg\"ue, in the Province of Mendoza, Argentina, has been taking
data stably since January 2004 while the construction was proceeding.
The construction was completed in mid 2008. Details of the design,
construction and performance of the Observatory can be found
in~\cite{paoEA,SDNIM,FDNIM}.

The Auger detector has been conceived with a cross-triggering
capability. Data are retrieved from both detectors whenever either
system is triggered\footnote{ Details about the triggers implemented
can be found in~\cite{SDTrAp} for the SD and in ~\cite{FDNIM} for the
FD.  }.  The surface array and the fluorescence telescopes allow the
reconstruction of extensive air showers with two independent
measurements. The combination of information from the two detection
subsystems enhances the reconstruction capability with respect to the
individual detector components~\cite{icrc2007BD,icrc2009CDG}. This
technique is called ``hybrid'' detection and the determination of the
exposure of the Observatory under this mode is the subject of the
present paper. The data period used for this purpose is between
November 2005 and May 2008. The exposure calculated here is the same
used for the energy spectrum measurement published
in~(\cite{AugerPLB2010}).

The paper is organized as follows. In section \ref{sec:hyb} we
describe the hybrid detection method. Section \ref{sec:method}
addresses the energy spectrum and the relevance of the hybrid exposure
to its determination. The effective data taking time in the hybrid
detection mode, i.e the hybrid on-time, and the different components
contributing to it are discussed in section \ref{sec:ontime}. The
Monte Carlo simulation used for the evaluation of the hybrid exposure
is described in section \ref{sec:sim}.  In section \ref{sec:vali} we
describe the event selection and make comparisons with data to
validate the Monte Carlo simulation. Finally in
section \ref{sec:exposure} we show the hybrid exposure as a function
of primary energy and in section \ref{sec:sum} we summarize.

\section{Hybrid data analysis}
\label{sec:hyb}
A {\it hybrid} event is an air shower that is simultaneously detected
by the fluorescence detector and the surface array.  If an air shower
independently triggers both detectors the event is tagged as a {\it
golden hybrid} and these events can be fully reconstructed in both
detection modes. In the SD the energy density of shower particles at
ground level is used to determine the cosmic ray energy. In the FD the
observation of the longitudinal profile of the shower allows the
measurement of the calorimetric energy of the primary particle. This
event sample, though small with respect to the SD sample, is very
important since it constitutes the base data set for the energy
calibration of the SD events \cite{SDspectrum,icrc2009CDG}.

The fluorescence detector, having a lower energy threshold, may
promote a sub-threshold trigger in the SD. In this case, surface
stations are matched by timing and location even though they do not
fulfil the conditions for an independent SD trigger. This is an
important capability because these sub-threshold {\it hybrid} events
would not have triggered the array otherwise. Here the energy
reconstruction relies uniquely on the calorimetric energy from the
longitudinal profile.

Like {\it golden hybrids} these events suffer statistical limitations,
but they are of particular interest because they allow an extension of
the measurement of the energy spectrum into a region where the SD is
not fully efficient~\cite{SDTrAp}. They have superior qualities with
respect to ``monocular'' FD events (those without SD information),
because of the precise measurement of the shower
geometry~\cite{FDNIM}.

In the FD, cosmic ray showers are detected as a sequence of triggered
pixels in a matrix of photomultipliers. This sequence allows the
determination of the shower-detector plane (SDP), the plane that
includes the location of the fluorescence detector and the line of the
shower axis, with a typical uncertainty of the order of a few tenths
of a degree.  Then the determination of the shower geometry relies on
the arrival times of photons in the individual pixels~\cite{FDNIM}.
In the monocular reconstruction the accuracy degrades when the
measured angular speed does not change significantly over the observed
track length. In such cases the shower axis can be largely
under-determined within the SDP, thus giving large uncertainties in
the reconstruction of the arrival direction and the impact point at
ground level. This further leads to uncertainties in other shower
parameters and in particular in the reconstructed shower energy.

\begin{figure}[t]  
\centering
   \includegraphics[width=0.95\textwidth]{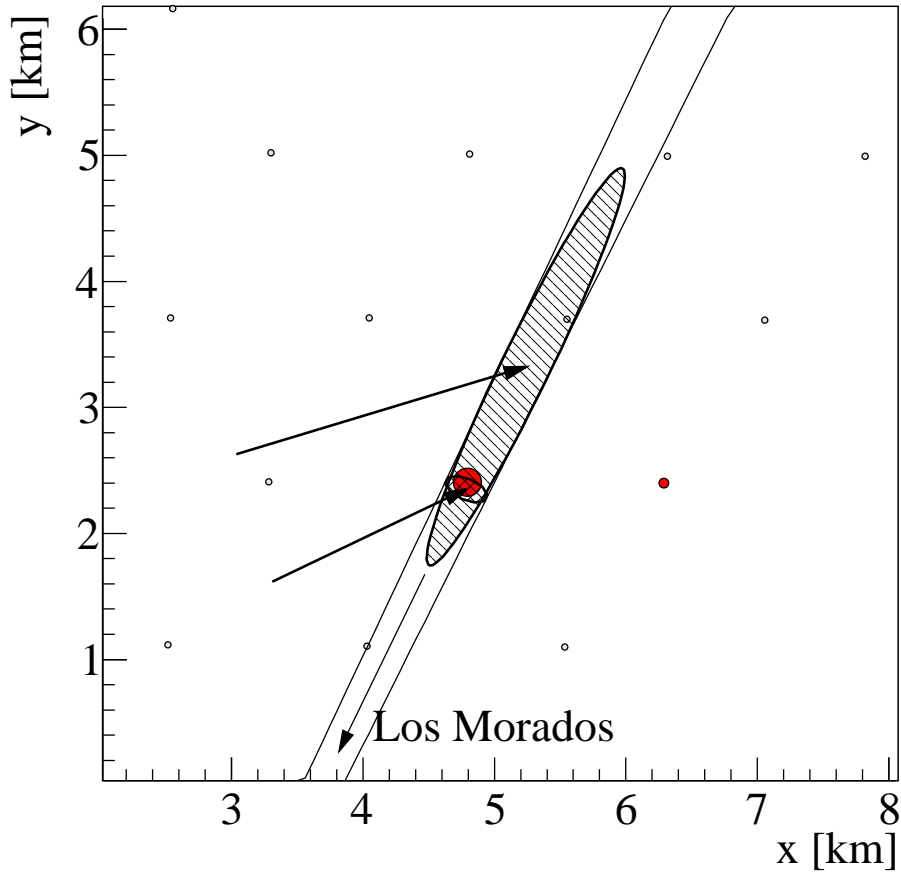} \caption{Determination
    of the impact point at the ground for a single event using both
    the mono and hybrid reconstruction methods.  The event has
    been detected by the Los Morados FD site: the downward-going arrow
    points towards the direction of the site and the two lines show
    the uncertainty of the SDP plane at ground level. The small (long)
    elongated ellipse represents the uncertainty on the core position
    in the hybrid (mono) reconstruction. The arrows indicate the
    reconstructed directions in the two cases, their length being
    proportional to the sine of the reconstructed zenith angle.  The
    open (full) circles show the active (triggered) SD stations.
    Triggered stations are shown with a radius proportional to the
    logarithm of the signal.  }  \label{fig:xyMonoHyb}
\end{figure}

The hybrid approach supplements the traditional FD direction fitting
method with the arrival time of the shower at the ground measured by a
single SD station. This results in a remarkable improvement in the
determination of the shower geometry, as illustrated in
figure~\ref{fig:xyMonoHyb} where the impact points at ground level
corresponding to mono and hybrid reconstruction methods are shown for
a typical event. Accurate knowledge of the shower arrival time at
ground level removes a degeneracy in the traditional FD monocular
approach that uses pixel timing to reconstruct the shower axis.  In
hybrid mode, the resolution of the direction and of the position of
the impact point at the ground are better than 0.6$^\circ$ and 50~m
respectively~\cite{BonifaziAng,Cris2006MM,icrc2007BD}.

The total energy of each event is obtained by combining the knowledge
of the detector response with monitoring data describing the
atmospheric conditions~\cite{FDNIM}.  Once the geometry is known, the
observed energy deposit profile is reconstructed taking into account
the scattering and the absorption of light during its propagation in
the atmosphere and the presence of forward-emitted and scattered
Cherenkov light. The method used is described in detail
in~\cite{unger_profile}. The energy released in the electromagnetic
part of the air shower is estimated by fitting a Gaisser-Hillas
function~\cite{ghfunc} to the reconstructed energy deposit profile and
integrating it over the entire range of atmospheric depth.  Finally,
the total energy of a shower is derived after correcting for the
invisible energy carried away by neutrinos and high energy
muons~\cite{Barbosa:2003dc}. After quality selection, the energy
resolution (defined as an event-to-event statistical uncertainty) of
the fluorescence detector is better than 10\%
\cite{icrc2007BD}. This value has been calculated from simulations and
has been cross checked with hybrid-stereo events, i.e. those events
which are detected and reconstructed in hybrid mode by more than one
FD-site\cite{FDNIM}. The energy resolution turned out to be energy
independent in the whole range.

Systematic uncertainties in the energy determination are related to
the detector, to the atmosphere and to the reconstruction
procedure. They are summarized in Tab.~\ref{tabhyb}. All these
uncertainties are found to be independent. A total uncertainty of
about 22$\%$~\cite{icrc2009CDG} is estimated by summing the individual
contributions in quadrature.

\begin{table}[t!]
\begin{center}
\caption{Current estimates of the systematic uncertainties affecting 
energy reconstruction. Values from~\cite{icrc2009CDG}.}
\vskip 0.3cm
\label{tabhyb}

\newcommand{\m}{\hphantom{$-$}}
\newcommand{\cc}[1]{\multicolumn{1}{c}{#1}}
\renewcommand{\tabcolsep}{0.3pc}
\renewcommand{\arraystretch}{1.3} 
\begin{tabular}{lclc}
\hline
     &  uncertainty \% &  & uncertainty \% \\ \hline  
  fluorescence yield (FY) &  14 & quenching effect on FY & 5\\
  FD absolute calibration & 9 & FD wavelength response & 3 \\ 
  molecular attenuation & 1 & aerosol attenuation & 7 \\
  multiple scattering model & 1 & FD reconstruction method & 10 \\
  invisible energy & 4 & &\\
\hline
\bf total && & \bf 22 \\
\hline
\end{tabular}\\[2pt]
\end{center}
\end{table}

\section{Energy Spectrum with hybrid events}
\label{sec:method}
The aperture of a cosmic ray instrument is per se a figure of merit of
its observation capability. The time integrated aperture is commonly
referred to as the exposure. In this section we discuss the relevance
of the exposure for the energy spectrum measurement. This is of
particular concern in the case of a detection based on fluorescence,
such as the hybrid case, where the time variations of the detection
and the inherent energy dependence make an accurate determination of
the exposure a key task.

The flux of cosmic rays $J$ as a function of energy is defined as:
\begin{equation}
  J(E)  =  \frac{\textrm{d}^{4}N_\mathrm{inc}}{\textrm{d}E \textrm{d}A \textrm{d}\Omega \textrm{d}t} \simeq \frac{\Delta N_\mathrm{sel}(E)}{\Delta E}\frac{ 1}{\mathcal E(E)};
  \label{eq.UHECRflux}
\end{equation}
where $N_\mathrm{inc}$ is the number of cosmic rays with energy
between $E$ and $E+$d$E$ incident on a surface element d$A$, within a
solid angle d$\Omega$ and time d$t$. $\Delta N_\mathrm{sel}(E)$ is the
number of detected events passing the selection criteria in the energy
bin centered around $E$, having width $\Delta$E. $\mathcal E(E)$
represents the energy-dependent exposure of the detector at the same
selection level.

The exposure, as a function of the energy of primary particle, can be
written as:
\begin{equation}
\mathcal{E}(E)  = \int_{T}\int_{\Omega}\int_{S_{gen}} \varepsilon(E,t,\theta,\phi,x,y) ~ \cos \theta ~ \textrm{d}S ~ \textrm{d}\Omega ~ \textrm{d}t = \int_{T} \mathcal A(E,t) ~ \textrm{d}t;
\label{eq.exposure}
\end{equation}
where $\varepsilon$ is the detection efficiency including the
different steps of the analysis, i.e trigger, reconstruction and
quality cuts, and $\textrm{d}S = \textrm{d}x \times \textrm{d}y$ is
the horizontal surface element.  $\textrm{d}\Omega
= \sin \theta \textrm{d}\theta \textrm{d}\phi$ and $\Omega$ are
respectively the differential and total solid angles.  The generation
area $\textrm{S}_\mathrm{gen}$ has been chosen large enough to exclude
any possible event detection and reconstruction outside it. $\mathcal
A(E,t)$ is the instantaneous aperture of the detector which depends on
the detector configuration at the time $t$.

The detector configurations of the Observatory have been continuously
changing over the period of data collection for the hybrid spectrum.
As construction of the SD progressed, the number of stations in
operation increased. Furthermore, even in a steady configuration, some
SD stations are temporarily out of service at any one time.  The SD
status is monitored by updating each second the list of ``active''
stations.  In principle the change in SD configuration is
straightforward to handle since the aperture is proportional to a
geometric area. In the case of a single missing SD station, the
effective area is slightly changed by about \unit[2]{km$^2$} at full
efficiency~\cite{SDTrAp}.

The FD detector configuration also changed with time during the
construction phase, with the number of telescopes changing from 12 to
24. In addition, a correction ring lens was added to each telescope
during the first two years of data taking. Thus, parts of the data
have been collected with different optical configurations. During
nightly operations individual telescopes are sometimes deactivated
because of increasing sky brightness, bad weather conditions or
hardware failures. Finally, the FD response is influenced by
atmospheric conditions such as the concentration of aerosols and cloud
coverage.

To properly take into account all the detector configurations and
their time variability a sample of events which reproduce the exact
conditions of the experiment (i.e its actual sequence of
configurations and on-time) has been simulated. This method, referred
to as {\it Time Dependent Detector Simulation}, is described in the
next sections. Given a set of $N$ simulated events generated on an
area $\textrm{S}_\mathrm{gen}$ within the time interval $T$, the
exposure eq.~(\ref{eq.exposure}) can be calculated numerically via
\begin{equation}
\mathcal{E}(E_\mathrm{rec})  = 
  2 \pi ~ \textrm{S}_\mathrm{gen} ~ \textrm{T} ~ 
   \sum_{i} \frac{n(E_\mathrm{rec}, \cos \theta_{i})}{N(E_\mathrm{gen},\cos \theta_{i})} ~ \cos \theta_{i} ~ \Delta \cos \theta_{i};
\label{eq.discrexposure}
\end{equation}
where $n$ denotes the number of events that fulfill the selection
criteria described in Sec.~\ref{sec:vali}. The exposure is calculated
as a function of reconstructed energy, $E_\mathrm{rec}$, to correct
for distortions of the steep energy 
spectrum due to the finite resolution of the energy reconstruction 
(see e.g.~\cite{Kalmykov:1969xt,Carvalho:2007mw} and Sec.~\ref{sec:Crosschecks}).

\section{Hybrid on-time}
\label{sec:ontime}
The efficiency of fluorescence and hybrid data taking is influenced by
many effects. These can be external, e.g. lightning or storms, or
internal to the data taking itself, e.g. DAQ failures. For the
determination of the {\it on-time} of the Pierre Auger Observatory in
the hybrid detection mode it is therefore crucial to take into account
all these occurrences and derive a solid description of the data
taking time sequence.

Data losses and inefficiencies can occur on different levels, from the
smallest unit of the FD, i.e. one single photomultiplier (pixel)
readout channel, up to the highest level, i.e. the combined SD-FD data
taking of the Observatory. To perform the time dependent detector
simulation we have to take into account all known disturbances and
then derive the on-time of the hybrid detection mode. To achieve this
aim we rely on a variety of monitoring information and the data set
itself. As a compromise between accuracy and stability we derived the
complete detector status down to the single pixel for time intervals
$T_{\mathrm{bin}} = 10~\mathrm{min}$.

\subsection{Telescope dependent sources}\label{telescope}

The active time of FD data acquisition is calculated using a minimum
bias data stream with a less restrictive trigger condition. This data
file includes sub-threshold FD events and is recorded at an event rate
about $8$ times higher than the standard rate of about $1$ event per
FD-site per $\mathrm{minute}$.

Even if the DAQ is running, the shutters of the telescope might be
closed due to bad weather alarms from the slow control system or other
failsafe mechanisms. To determine the status of the shutters we use
the information on night sky background level provided by algorithms
implemented in the front-end electronics boards. Every 30\,s data from
each PMT channel is written to a monitoring data file which records
parameters including ADC-variance, baseline, First Level Trigger (FLT)
threshold and trigger frequency for each pixel~\cite{FDNIM}.

\begin{figure}[!t]
\begin{center}
  \includegraphics[width=0.95\textwidth]{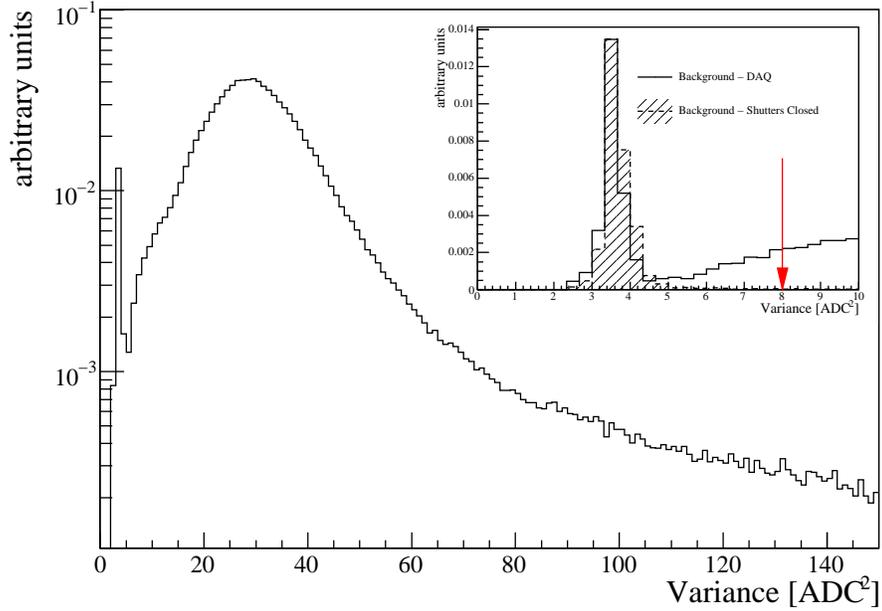}
\caption[]{Distribution of background variances. The main contribution to 
this background noise is the night-sky background light coming from
stars and the direct and scattered moonlight. The upper-right panel
shows a magnified view of the low variance region superimposed on data
recorded with closed shutters (shaded histogram). The arrow shows the
variance threshold used to select good data.}
\label{fig:BGVariances}
\end{center}
 \end{figure} 

The ADC-variance distribution from these data is shown in
figure~\ref{fig:BGVariances}. Background data is also collected during
the nightly relative calibration runs, i.e. with closed shutters (see
the upper-right panel in figure~\ref{fig:BGVariances}).  A mean value
of about $3.5~\mathrm{ADC}^2$ is obtained in these
conditions\footnote{ Muons hitting the pixel camera is the main source
of the noise triggers. }.  For each time interval the efficiency of
open shutters is then derived as:
\begin{center}
 \begin{equation}
  \varepsilon_{\mathrm{shutter}}= \frac{T_{\mathrm{open}}}{T_{\mathrm{bin}}}
 \end{equation}
\end{center}
where $T_{\mathrm{open}}$ denotes the time (for a given telescope) for
which the mean variance over the whole camera is larger than
$8~\mathrm{ADC}^2$ .  If background data are not available, no
efficiency is calculated. The status flag $\delta_{\mathrm{tel}}$ is
then set to 0.

The deadtime due to the finite readout speed of the DAQ system must
also be taken into account.  The deadtime is stored on an
event-by-event basis in the output of the FD data acquisition. For
each telescope, this deadtime $T_{\mathrm{DAQ}}^{\mathrm{dead}}$ is
converted into an efficiency of detecting cosmic ray data in the given
time interval by:

\begin{center}
 \begin{equation}
  \varepsilon_{\mathrm{DAQ}}= 1 - \frac{T_{\mathrm{DAQ}}^{\mathrm{dead}}}{T_{\mathrm{DAQ}}}
 \end{equation}
\end{center}

where $T_{\mathrm{DAQ}}$ is the total running time of the DAQ in the given time
interval.

\subsection{FD-site dependent sources}\label{eye}
Currently two possible sources of inefficiency are known to affect the
data taking at the FD-site level.

The first is due to the atmospheric monitoring system. An FD veto is
set by the Lidar system before performing laser shots in the field of
view of a fluorescence detector. The cumulative Lidar veto time is
stored on an event-by-event basis in the data files. This deadtime
$T_{\mathrm{Lidar}}^{\mathrm{dead}}$ is converted into an efficiency
by:

\begin{center}
 \begin{equation}
  \varepsilon_{\mathrm{Lidar}}= 1 - \frac{T_{\mathrm{Lidar}}^{\mathrm{dead}}}{T_{\mathrm{DAQ}}}
 \end{equation}
\end{center}

This efficiency can be interpreted as the probability of a cosmic ray
event falling outside the Lidar vetoed period.

To extend the hybrid detection capability below the SD trigger
threshold~\cite{SDTrAp,SDSpectrumPRL}, all the FD triggers are sent to
and processed by the central data acquisition system (CDAS). It reads
out the portion of the surface array closest to the relevant
fluorescence building. Then FD and SD data streams are merged to form
hybrid events.  A source of inefficiency comes from the protection
algorithm implemented in the CDAS to prevent the acquisition of long
periods of excessive event rates\footnote{Lightning and other noise
events may cause higher FD trigger rates which would cause significant
deadtime for the surface array due to the finite readout time of the
array.}. This veto mechanism induces the loss of hybrid events.  An
estimate of the event loss probability in a given time interval is
calculated by comparing events from the FD data files and from the
final merged hybrid files (which only include those sent to CDAS).
This recovery mechanism is energy dependent as it is related to the SD
trigger probability~\cite{SDAperture_ICRC2005} and is accounted for on
an average basis. $\langle \varepsilon_{\mathrm{T3veto}}(s,t) \rangle$
is the resulting average efficiency for each FD site {\it s} and time
{\it t}.

\subsection{CDAS status}\label{cdas}

CDAS inefficiencies must also be taken into account.  The surface
detector array is constantly monitored and a very detailed description
of the array status is available with a time resolution of 1
second. In addition to the usually very localized problems of single
SD stations, time periods with trigger related problems~\cite{SDTrAp}
are excluded in the hybrid on-time via the CDAS status flag
$\delta_{\mathrm{CDAS}}$. Given a constant rate of hybrid events
$\lambda$, the probability $P$ that the time interval between two
consecutive hybrid events is larger than $T$ is given by $P(T) =
e^{-\lambda T}$. Taking $\lambda \approx 1.7 \times 10^{-2}$ Hz (1
event per minute) and $T=600$ sec, then $P = 3.7 \times 10^{-5}$.
Based on this calculation, an additional check is performed requiring
at least one hybrid event per $\unit[10]{min}$ time interval.

\subsection{Results and cross-checks}\label{otres}

For each time $t$ in a given time slot of duration 
T$_{\mathrm{bin}}$, the fraction of
operational time $f(i,t)$, for the telescope $i$ belonging to the FD
site $s$, can be written as:

\begin{eqnarray}
  f(i,t) &=& \varepsilon_{\mathrm{shutter}}(i,t) \cdot  \varepsilon_{\mathrm{DAQ}}(i,t)
  \cdot \delta_{\mathrm{tel}}(i,t) \\
     \nonumber & & \cdot \varepsilon_{\mathrm{Lidar}}(s,t) \cdot
  \langle \varepsilon_{\mathrm{T3veto}}(s,t) \rangle \\
   \nonumber & &\cdot \delta_{\mathrm{CDAS}}(t) 
\end{eqnarray}

where the $\varepsilon$'s identify the efficiencies due to the
different sources and the $\delta$'s are status flags
($\delta=[0,1]$).  All the expected sources of inefficiencies have
been described in detail in the previous sections.

\begin{figure}[!t]
\begin{center}
  \includegraphics[width=0.95\textwidth]{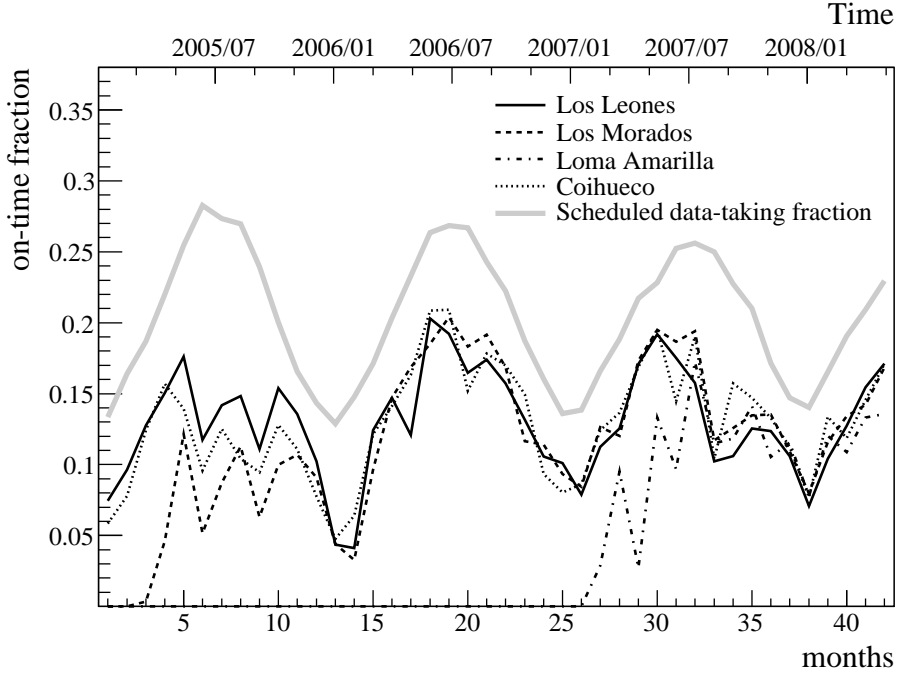}
\caption[]{Time evolution of the average hybrid on-time fraction 
           during the construction phase of the Pierre Auger Observatory.
           Both the seasonal modulation and the starting of 
           commissioning phases of the different FD-sites are visible.  
           Gray line represents the scheduled data-taking time fraction 
           limited to the nights with moon-fraction lower than 60\%.}  
           \label{fig:OnTime}
\end{center}
\end{figure}

The time evolution of the full hybrid duty-cycle over 3 years during
the construction phase of the observatory is shown in
figure~\ref{fig:OnTime}. It shows the on-time fraction, defined as the
ratio of the overall on-time to the time duration of each interval.
To avoid pile-up effects in the plot, time bins are chosen to coincide
with FD data-taking shifts. Data-taking is currently limited to dark
periods with moon-fractions smaller than 60\% as seen by each
individual telescope: this leads to about 16 nights of data taking per
moon-cycle. The scheduled data-taking time fraction is also shown in
figure 3 (gray line). A seasonal modulation is clearly visible, since
higher fractions are observed in the austral winter during which the
nights are longer.

Note that the FD-site at Los Morados became operational in May 2005
and that at Loma Amarilla started in March 2007. After the initial
phase of commissioning, the mean on-time is about 12\% for all
FD-sites, which corresponds roughly to about 70\% of the scheduled
time fraction. This efficiency is primarily due to weather effects
with a minor part determined by detector effects.

A validation of the on-time determination and an estimate of its
systematic uncertainty has been performed using data from the Central
Laser Facility (CLF)~\cite{CLF}. These data are embedded in the
standard FD data stream.  As CLF laser shots can be observed from all
FD-sites, one can calculate the conditional probability of recording
the laser signal in a particular site $s$ given at least one other
observation in any other site.  The expected number of laser shots in
site $s$ can be derived from the on-time of the telescope pointing to
the CLF.  The laser observation probability is obviously dependent on
the transmission coefficient of the atmosphere.  The probability of
observing a laser during aerosol-free periods, i.e with vertical
aerosol optical depth VAOD $\approx$ 0, is expected to be 100\%. A
small deviation from this value of about 4\% was found and the on-time
has been corrected accordingly to account for possibly lost periods.

\section{Monte Carlo Simulation}
\label{sec:sim}
For the calculation of the hybrid exposure, the size of the simulated
event sample is crucial for acceptable statistical and systematic
uncertainties. For this purpose the simulation activity followed a
graded approach with full Monte Carlo analysis for specific studies,
like the trigger efficiency, and fast simulations, validated with the
full Monte Carlo method, when high statistics were required.

\subsection{Trigger efficiency}
\label{sim:TriggerEff}
A complete Monte Carlo hybrid simulation has been performed to study
the trigger efficiency and the detector performance.  The simulation
sample consists of about 6000 proton and 3000 iron
CORSIKA~\cite{corsika} showers with energies ranging between 10$^{17}$
and \energy{19.5}. These energies are of particular interest for the
trigger studies since they cover both SD and hybrid thresholds. The
showers have been generated using respectively
QGSJET-II~\cite{qgsjet,qgsjetII} and FLUKA\cite{fluka} as high and low
energy hadronic interaction models.  The FD simulation
chain~\cite{FDsim} reproduces in detail all the physical processes
involved in the fluorescence technique. It includes the generation of
fluorescence and Cherenkov photons in the atmosphere, their
propagation through the atmosphere to the telescope aperture, the
ray-tracing of photons in the Schmidt optics of the telescopes, and
the simulation of the response of the electronics and of the
multi-level trigger.  The surface detector response is simulated using
Geant4~\cite{geant4} within the framework provided by the Auger
Offline software~\cite{Offline}.  For this particular purpose we
assume the SD array is fully operational and deployed.

\begin{figure}[t]  
  \begin{center}
  \vskip -0.3 cm
    \includegraphics[width=0.95\textwidth]{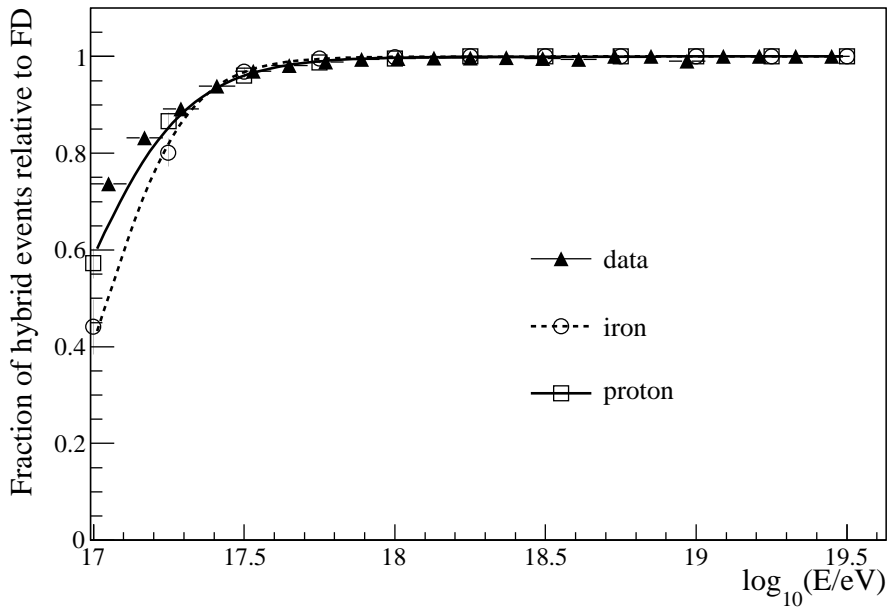}
    \caption{Relative hybrid trigger efficiency from hybrid simulation
      for proton and iron primaries. The hybrid trigger efficiency
      calculated using data is also shown.}
    \label{fig:fullhyb}
  \end{center}
\end{figure}

In figure~\ref{fig:fullhyb} it is shown the hybrid trigger
efficiency, i.e. the probability of detecting a fluorescence event in
coincidence with at least one triggered SD station, is flat and equal
to 1 at energies greater than \energy{18}, independent of primary
mass.
The difference between proton and iron primaries increases at lower
energies but is negligible at energies as low as
\energy{17.5}. Protons are slightly more efficient than iron primaries at the
lowest energies.  This is mainly due to the larger fraction of proton
events interacting deeper in the atmosphere.  The hybrid trigger
efficiency from fluorescence data is also shown in
figure~\ref{fig:fullhyb}. Only events landing on an active part of the
surface detector have been selected and minimal quality cuts have been
applied in order to have a reliable reconstructed energy and to safely
derive the trigger probability curve.  Data and simulation
consistently show that a fluorescence event is always hybrid for
energies larger than \energy{18}.

In addition, the probability of a shower triggering a given SD station
has been studied as a function of primary cosmic ray energy, mass,
direction and distance to the shower axis, and a set of ``Lateral
Trigger Probability'' (LTP) functions have been derived and
parameterised~\cite{ltppaper}.  For a vertical proton primary shower,
each station is on average fully efficient within a distance of 750,
1000, 1300, and 1600 m at energies
of \energy{17.5}, \energy{18}, \energy{18.5} and \energy{19},
respectively.  Details on this study are discussed in \cite{ltppaper}.

\subsection{Fast simulation}
\label{sim:Conex}

To follow and reproduce the time dependence of the hybrid exposure,
each detector configuration must be taken into account. This
approach requires a large number of simulations. The method used to
achieve this goal within a reasonable computational time relies on the
simulation of longitudinal shower profiles generated with
CONEX~\cite{conex}, a fast generator based on CORSIKA \cite{corsika}
shower code. After the simulation of the first few ultra-high
energy interactions, CONEX switches to numerical solutions of the
underlying cascade equations that describe the evolution of the
different shower components. Although this method is extremely fast,
the most important features provided by full Monte Carlo simulations,
including shower to shower fluctuations, are very well
reproduced~\cite{conex,Pierog_ICRC07}.

The simulation of the FD response proceeds as in the full method
discussed above. Since no ground level particles are generated by
CONEX, the SD response cannot be directly simulated. In this case the
SD trigger is reproduced using the LTP parameterisation functions.
The actual status of the SD array is retrieved using the time of each
simulated event. The event trigger probability is then calculated as
the convolution of all the LTPs of the working SD stations. This is
particularly important for low energy and inclined events.

The SD timing information needed in the hybrid reconstruction mode is
provided by a simplified simulation (i.e. SdSimpleSim) implemented in
the Offline simulation framework.  With this approach the lateral
distribution of the air shower is assumed to follow a NKG-like
functional form~\cite{nkgfuncnk,greisen_1965}.  A model generating
realistic signal timing for the closest station to the shower axis has
been derived from a full Monte Carlo using AIRES~\cite{aires}
simulations.  The SdSimpleSim code also includes the simulation of
noise triggered stations, which could spoil the reconstruction of the
event.  The noise rate of the surface detector is self-adjusting to
yield $\unit[20]{Hz}$ per station. As a cross-check, the number of
noise triggered stations has been derived from data and the obtained
distributions have been parameterized.

\begin{figure}[t]  
 \centering   
   \subfloat[]{\label{fig:SdEyeTimeDistrib}
\includegraphics[width=0.51\textwidth]{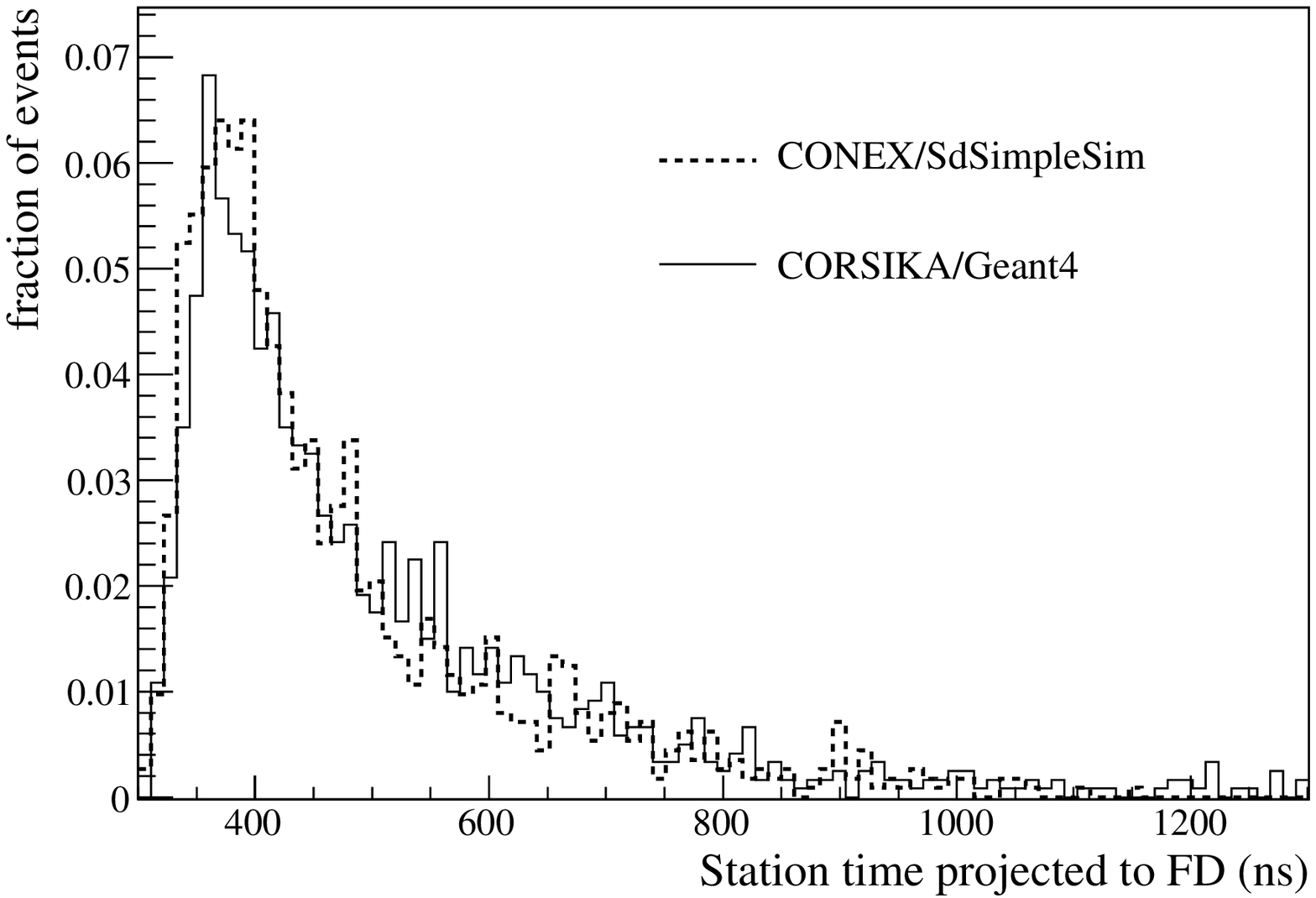}}
   \subfloat[]{\label{fig:ResCONEXvsCORSIKA}
\includegraphics[width=0.51\textwidth]{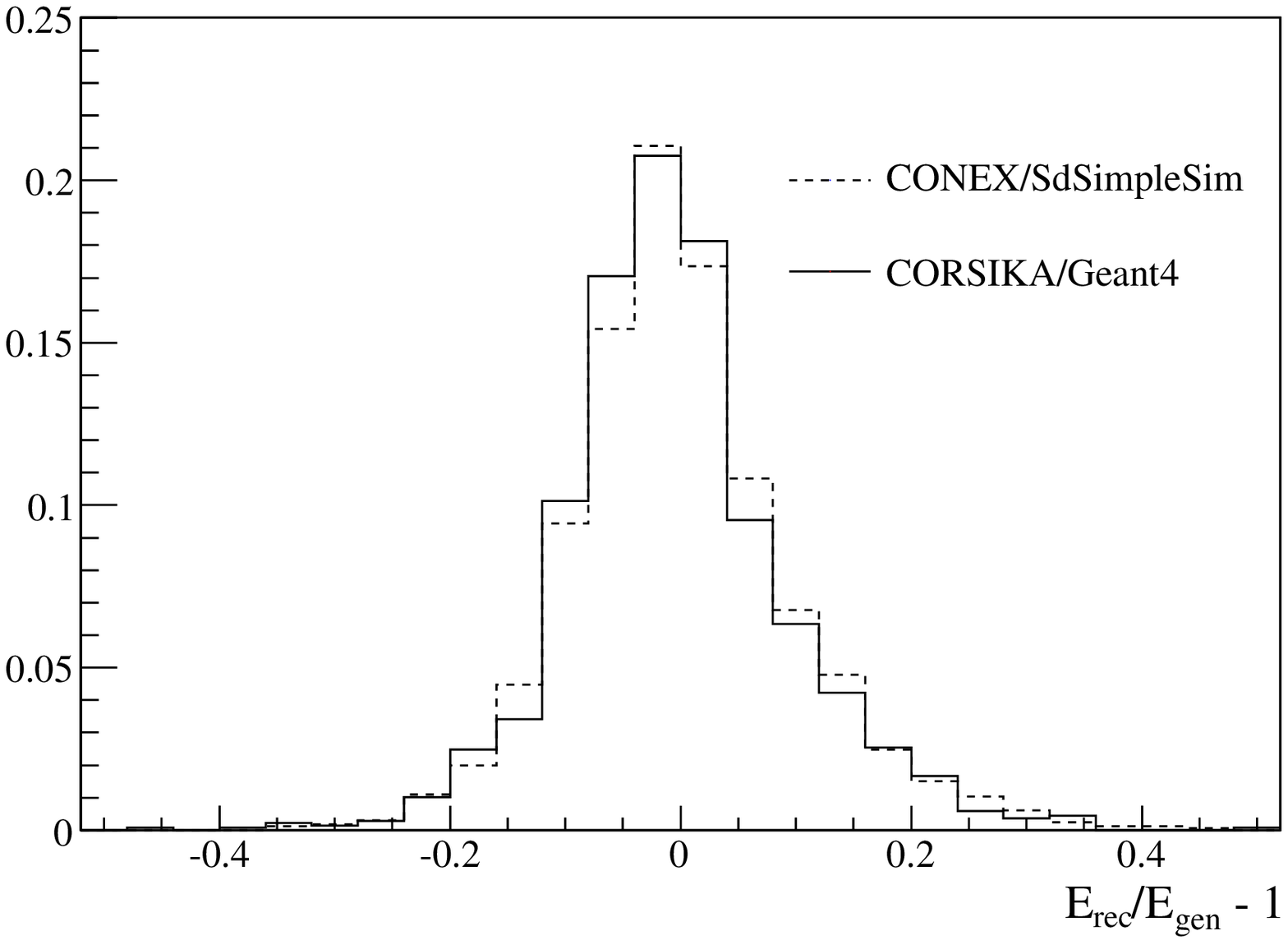}} 
    \caption[]{Comparison between CORSIKA/Geant4 simulations and the
        fast CONEX/SdSimpleSim approach. (a): distribution of the time
        at which the SD station is triggered. (b): difference between
        the simulated and the reconstructed energies using the hybrid
        technique. The figures refer to events at log$_{10}$(E/eV) =
        $18.5$.}  \label{fig:time_ene}
\end{figure}

Dedicated CORSIKA/Geant4 simulations have been carried out to validate
the performance of this fast approach against the full Monte Carlo
method.  Figure~\ref{fig:time_ene} shows the distribution of the
station trigger times and the difference between simulated and
reconstructed energy as obtained with the two simulation modes. The
consistency between these results provides a robust validation of the
fast approach and makes it possible to produce of huge number of
simulated events.

\subsection{Time Dependent Detector Simulation}
\label{sim:TDDS}

The Monte Carlo simulation for the calculation of the hybrid exposure
has been based on the fast simulation approach described above. In
fact for covering all the energy ranges and the phase space of the
detector configurations with enough statistical power, the number of
simulated events is required to be largely oversampled with respect to
the available raw data. The simulation has been designed to reproduce
the actual sequencing of the detector status with a resolution of
\unit[10]{min} which corresponds to the time bin slot used for the
on-time calculation.  First a time is randomly chosen within the
sidereal time interval we want to simulate. Then all the relevant
status information about each detector is retrieved from the on-time
calculations.  Based on the on-time fraction during the simulated time
bin, only a sub-sample of the events is sent to the detector
simulation.

The CONEX showers used for this purpose have been generated from
$10^{17}$ up to \energy{21}. QGSJET-II~\cite{qgsjet,qgsjetII} and
Sibyll~\cite{Ahn:2009wx} have been used as high energy interaction models.
Proton and iron particles are taken as cosmic ray primaries.

To account for the growth of the array with time and problems during
the SD data-taking, only the active SD stations are considered during
simulation.

For the FD time dependent simulations the values of variance, baseline
and trigger threshold averaged over \unit[10]{min} are considered.
The available FD absolute calibration data are used to adjust the
simulated electronic gains on a pixel by pixel basis. This scales the
shower signal with respect to the FADC trace noise and therefore
influences the signal-to-noise ratio. In addition, incorrect cabling
in some FD cameras is simulated for the instances discovered in the
real detector. Data from the atmospheric monitoring system is used to
set the hourly aerosol density profile as measured by the
CLF~\cite{CLF} and the monthly mean molecular atmosphere as provided
by balloon flights~\cite{FDAtm}.

\section{Event selection and validation of Monte Carlo simulation}
\label{sec:vali}
An unbiased measurement of the cosmic ray flux requires an exposure as
free as possible from systematics. To this aim only high quality
hybrid events are used.

\subsection{Quality cuts}
\label{sec:quality}

In this analysis high quality hybrid events have been selected using
the following criteria:
\begin{itemize}
\item since we use the Gaisser-Hillas function~\cite{ghfunc} to
  evaluate the total calorimetric energy, a successful fit of the
  longitudinal profile with this function is required.  Moreover, the
  $\chi^{2}$ per degree of freedom of the fitted profile should be
  less than 2.5;
\item the energy and shower maximum can only be reliably measured if
  X$_{max}$ is in the field of view (FOV) of the telescopes (covering
  $1.5^\circ$ to $30^\circ$ in elevation).  Events for which only the
  rising or falling edge of the profile is detected are not used, i.e.
  it is required that the depth of shower maximum be within the
  minimum and maximum observed depths;
\item to avoid potential systematic uncertainties related to the
  calculation of the Cherenkov light contribution, events with a
  relative amount of reconstructed Cherenkov light exceeding $50\%$ of
  the total received light are not used in this analysis;
\item a good energy resolution is assured by accepting only events for
  which the total uncertainty of the reconstructed energy (including
  the propagated statistical uncertainties of the detected photons,
  the geometry and the atmosphere) is smaller than 20\%.
\end{itemize} 
Furthermore it is required that:
\begin{itemize}
\item the aerosol content of the atmosphere is 
  measured~\cite{CLF,Lidar} for the time period of the event to allow
  a precise calculation of the transmission and scattering of photons
  in the atmosphere;
\item the cloud coverage according to Lidar measurements~\cite{Lidar}
  is lower than $25\%$ at the time of the event, since clouds could
  obscure part of the longitudinal profile and lead to an event
  selection inefficiency not accounted for in the aperture simulation.
\end{itemize}

\subsection{Fiducial cuts}
\label{sec:fiducial}

 \begin{figure}[!t]
   \centering   
   \subfloat[quality selection]{\label{fig:Exposure_PrimaryDifference_Quality}
     \includegraphics[width=0.51\textwidth]{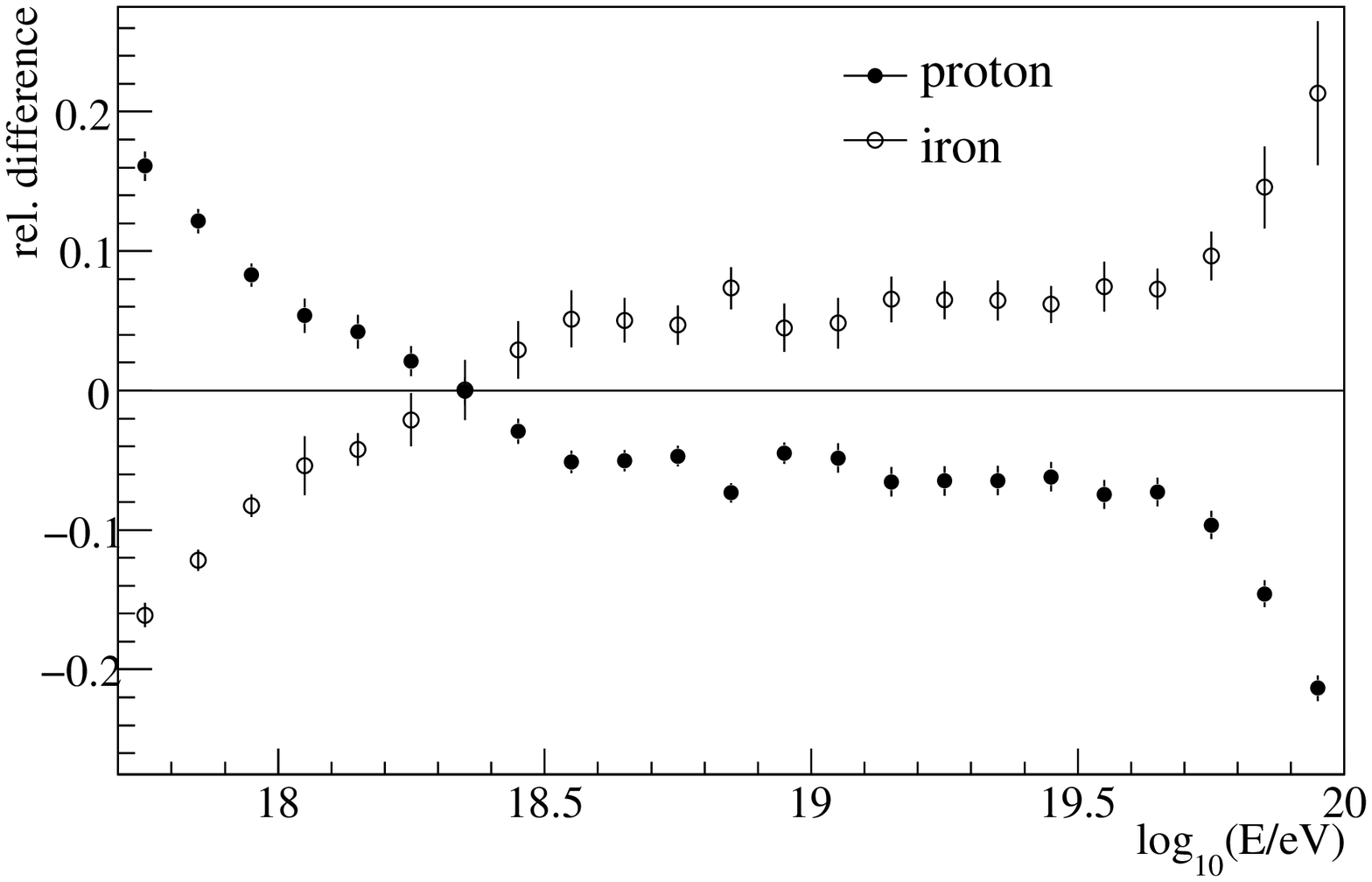}}
   \subfloat[quality and fiducial volume]{\label{fig:Exposure_PrimaryDifference_Fiducial}
     \includegraphics[width=0.51\textwidth]{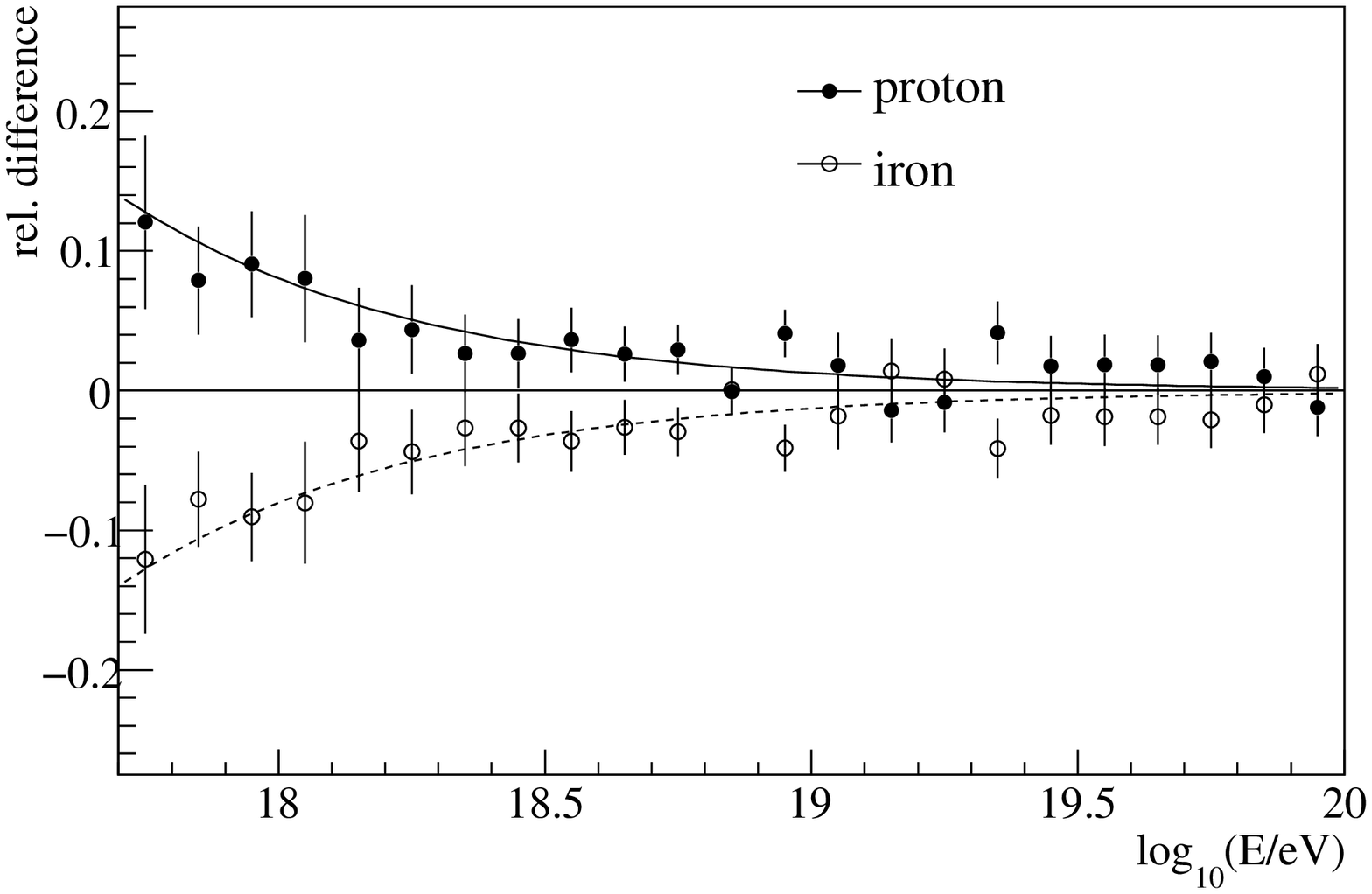}}
   \caption[]{Relative difference between proton and iron exposure
     with respect to a mean composition exposure, as obtained from the
     Time Dependent Detector Simulation. In the left panel
     the dependence on the primary mass is clearly visible. In the
     right panel this difference is strongly reduced by the fiducial
     volume cut.}
   \label{fig:primary_difference}
 \end{figure}

In addition to the above mentioned quality criteria, fiducial cuts
have been applied to assure a robust calculation of the exposure,
independent of the trigger threshold, mass composition and energy
scale uncertainty.

To ensure that the probability of a trigger from at least one surface
detector station is unity regardless of the primary particle, it is
required that
\begin{itemize}
\item the energy of the shower is larger than \energy{18};
\item the zenith angle of the shower is less than 60$^\circ$;
\item the position of the station used for the hybrid reconstruction
  is within 1500~m of the shower axis.
\end{itemize}

The limited field of view of the fluorescence detector and the
requirement of observing the shower maximum may both introduce a
different selection efficiency for different primary masses. For
instance, protons develop deeper into the atmosphere and have a deeper
shower maximum than heavy primaries, on average. For vertical events
the fraction of events with their maxima falling below the observation
level is thus larger for proton primaries and correspondingly the
selection efficiency is smaller.  The mass dependence of the exposure
for showers selected only by quality cuts is clearly visible in
figure~\ref{fig:Exposure_PrimaryDifference_Quality}.  At low energies,
where events are only detected close to the detector, iron primaries
have a smaller exposure because of their shallower \Xmax that tends to
be more often above the upper limit of the FD field of view than it is
for protons. At high energies the majority of the showers are far away
from the telescopes and the bias is dominated by the lower field of
view boundary that disfavors the selection of primary protons.

In order to achieve an almost equal detection probability for all
possible primaries, the following fiducial {\it field of view} cut has
been designed:

\begin{eqnarray}
 X_{\textrm{up}} \;[\depth{}]   & \geq  &  900 + 6\cdot( \varepsilon-18) \\
 X_{\textrm{low}}\;[\depth{}]   & \leq  & \left\{
 \begin{array}{ll}
  550 - 61 \cdot (\varepsilon - 19.06)^2  &\qquad \mathrm{for}\quad  \varepsilon < 19.06 \\
  550  &\qquad \mathrm{for} \quad \varepsilon  \geq 19.06 
 \end{array}
 \right.
 \end{eqnarray}

where $\varepsilon = \log_{10} (E/\!\eV)$, and X$_{\textrm{up}}$ and
X$_{\textrm{low}}$ are the upper and lower boundaries of the telescope
field of view which depend on the shower geometry.  The application of
this cut reduces the primary mass dependence above \energy{18}. This
is shown in figure~\ref{fig:Exposure_PrimaryDifference_Fiducial}.

 \begin{figure}[!t]
   \centering
   \subfloat[]{\label{fig:TriggerEfficiency_EnergyScale}%
     \includegraphics[width=0.52\textwidth]{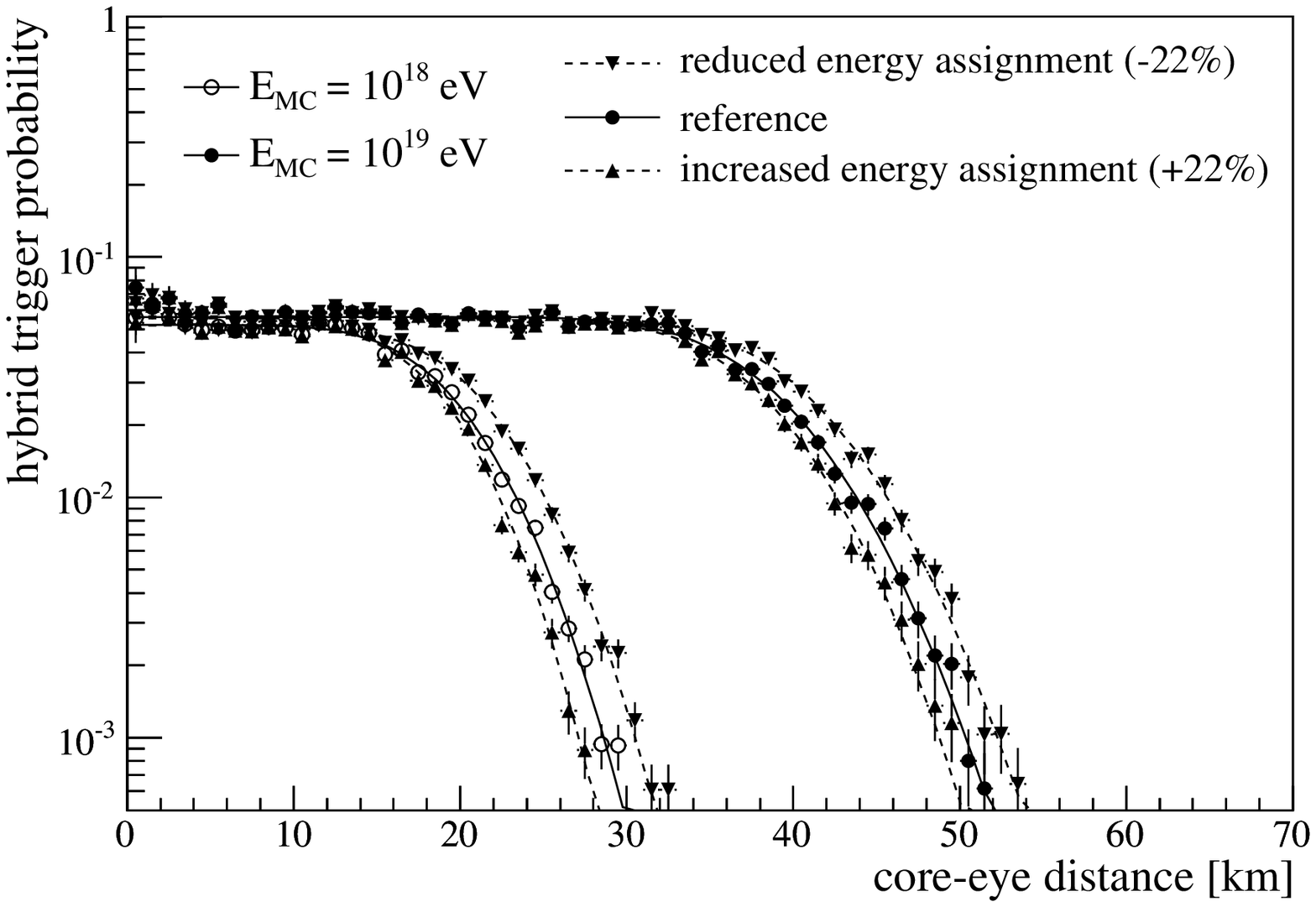}}
   \subfloat[]{\label{fig:TriggerEfficiency}%
     \includegraphics[width=0.52\textwidth]{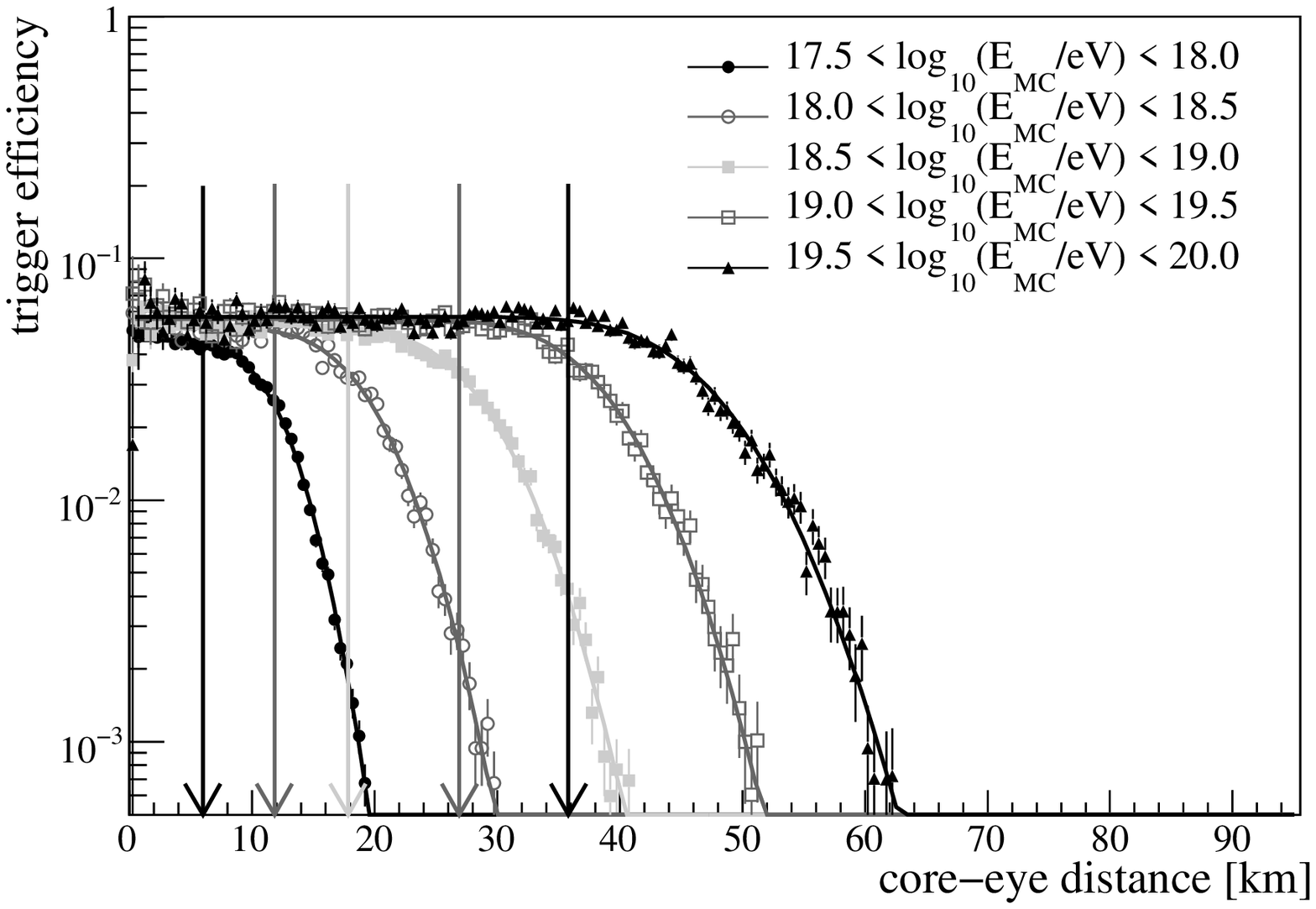}}

  \caption[fiducial distance selection]{The systematic uncertainty of
    the absolute energy scale of about $\unit[22]{\%}$ would cause
    significant systematic uncertainties in the fiducial volume and
    the exposure (left panel). Dedicated event
    selection criteria are used to remove this dependency
    (right panel). The arrows show the cut values
    for the different energies based on
    eq. (\ref{fig:fiducialVolume}).}
   \label{fig:energy_scale_uncertanty}
 \end{figure}

A further cut was introduced in order to remove the FD trigger
threshold effects induced by the energy scale uncertainties.  The
fluorescence detector trigger is in fact fully efficient for short
distances between the shower and the detector. At larger distances the
trigger probability decreases. A possible systematic shift in the
assignment of the shower primary energy, due to the 22\% energy scale
uncertainty (see tab.~\ref{tabhyb}), may alter the derived trigger
threshold and the exposure. To quantify this effect, the energy
assignment of the simulated events can be shifted up and down by 22\%
before applying the selection criteria. The effect is shown in
figure~\ref{fig:TriggerEfficiency_EnergyScale}.

The possible dependence of the trigger threshold on a systematic shift in the
energy assignment has been removed by dedicated selection criteria
obtained from Monte Carlo studies.  The available detection volume is
limited by a set of fiducial volume cuts which require the shower core
to lie within a distance $D_\textrm{max}$ from the fluorescence
detectors:

\begin{eqnarray}
D_\textrm{max}\;[\textrm{km}] & \leq &\left\{
\begin{array}{ll}
  24 + 12 (\varepsilon - 19)  \;\;\hfill \qquad  \mathrm{for}\quad  \varepsilon  < 18.5 \\
  24 + 12 (\varepsilon - 19) + 6(\varepsilon - 18.5)   \;\;\hfill \qquad \mathrm{for}\quad  \varepsilon   \geq 18.5 
 \end{array}
\right.
\label{fig:fiducialVolume}
\end{eqnarray}

As is clearly shown in figure~\ref{fig:TriggerEfficiency} this cut
limits the available detection volume to a region in which the
fluorescence trigger is saturated even if the energy scale is changed
within the known systematic uncertainties ($\pm 22 \%$).
The exposure calculation thus becomes independent of the trigger
threshold and of the absolute energy scale within current estimations
of its systematic uncertainty.

   \begin{figure}[!t]
   \centering   
   \subfloat[core-station distance]{\label{fig:CoreTankDistance}%
     \includegraphics[width=0.52\textwidth]{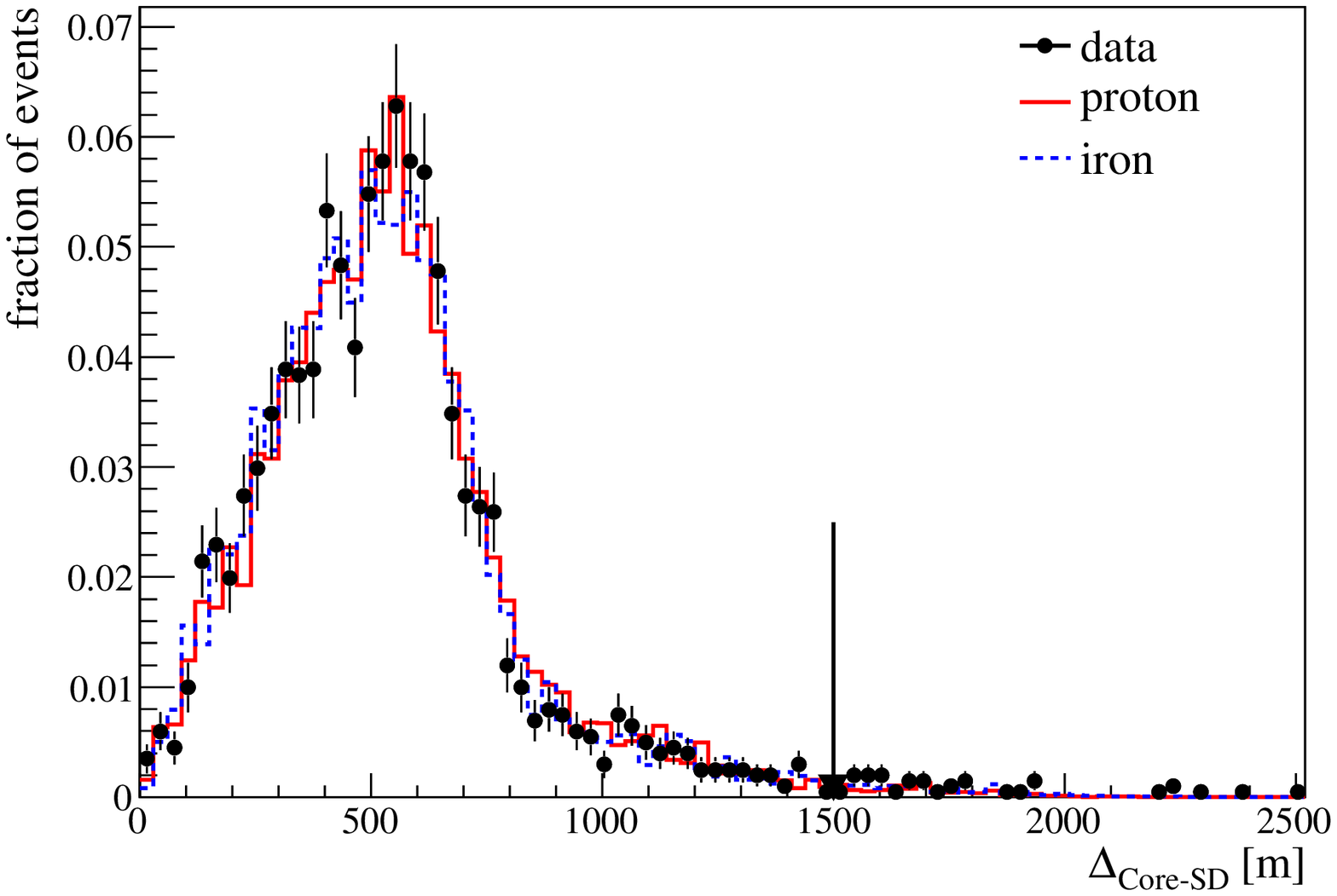}}%
   \subfloat[profile $\chi^{2}/ndof$ for the profile fit]{\label{fig:MaxHole}%
     \includegraphics[width=0.52\textwidth]{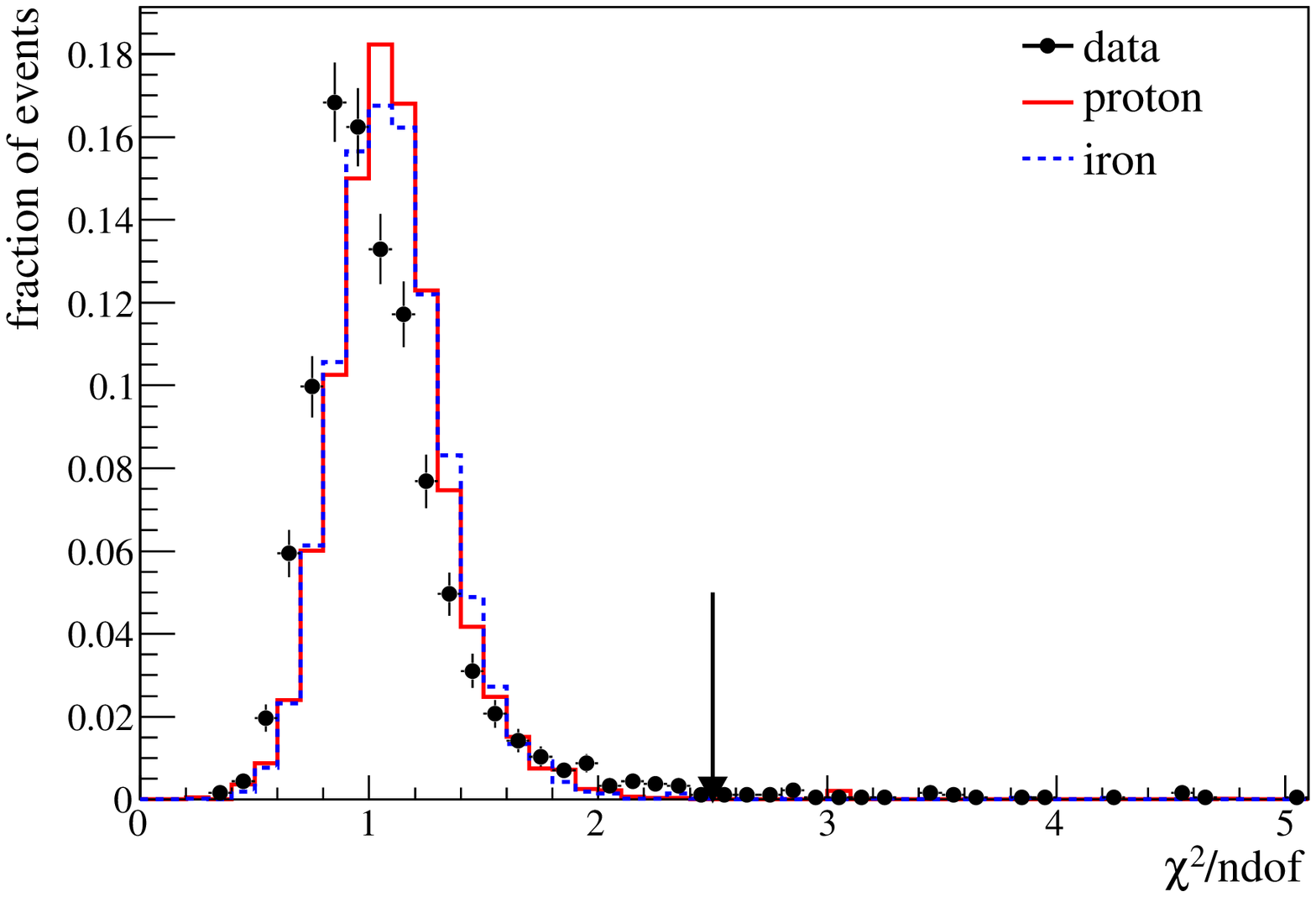}}
   \quad
    \subfloat[zenith angle]{\label{fig:Zenith}
     \includegraphics[width=0.52\textwidth]{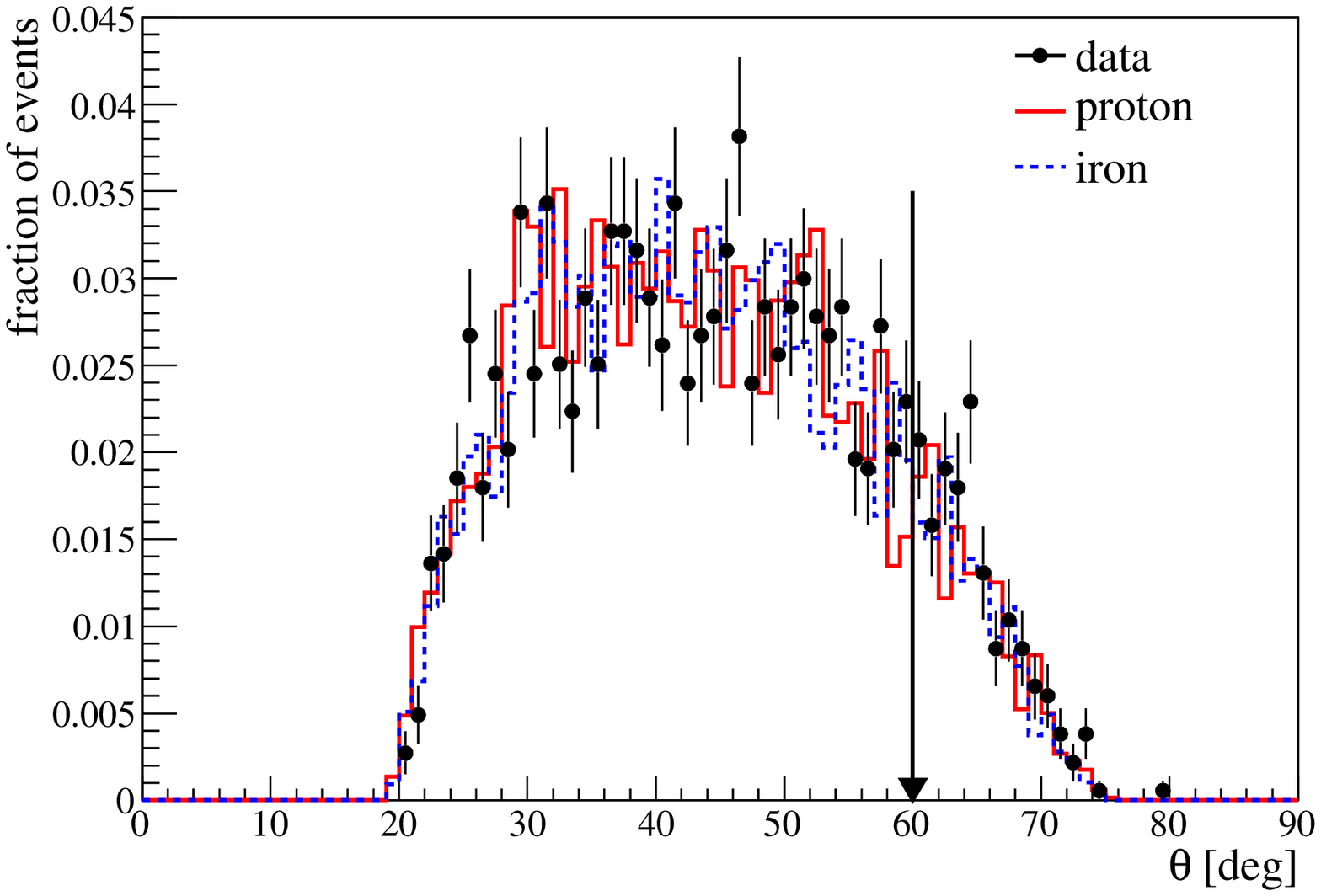}}
   \subfloat[Cherenkov fraction]{\label{fig:CherenkovFraction}
     \includegraphics[width=0.52\textwidth]{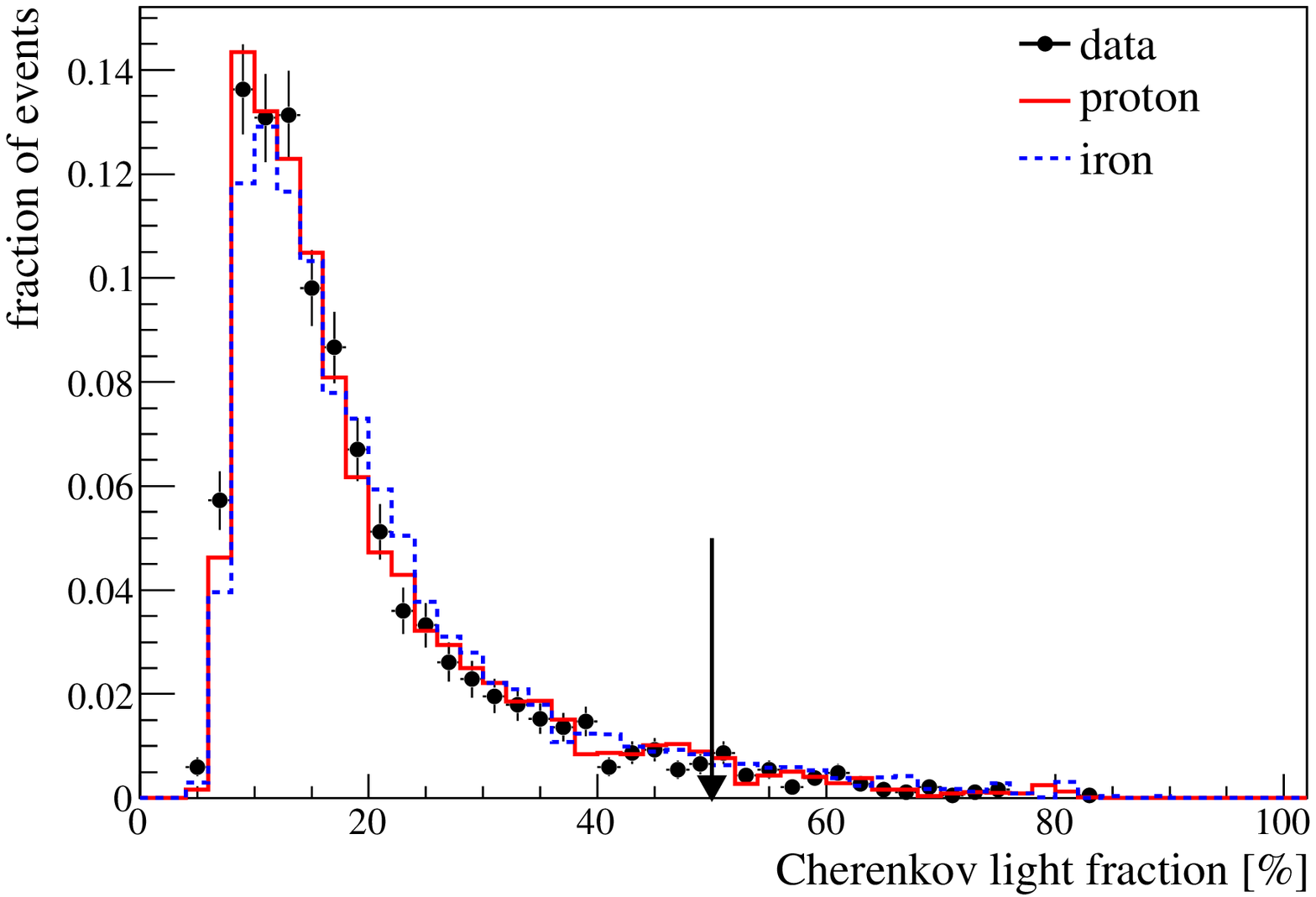}}
   \quad       
   \subfloat[distance of X$_{max}$ to FOV boundary]{\label{fig:XmaxFOV}
     \includegraphics[width=0.52\textwidth]{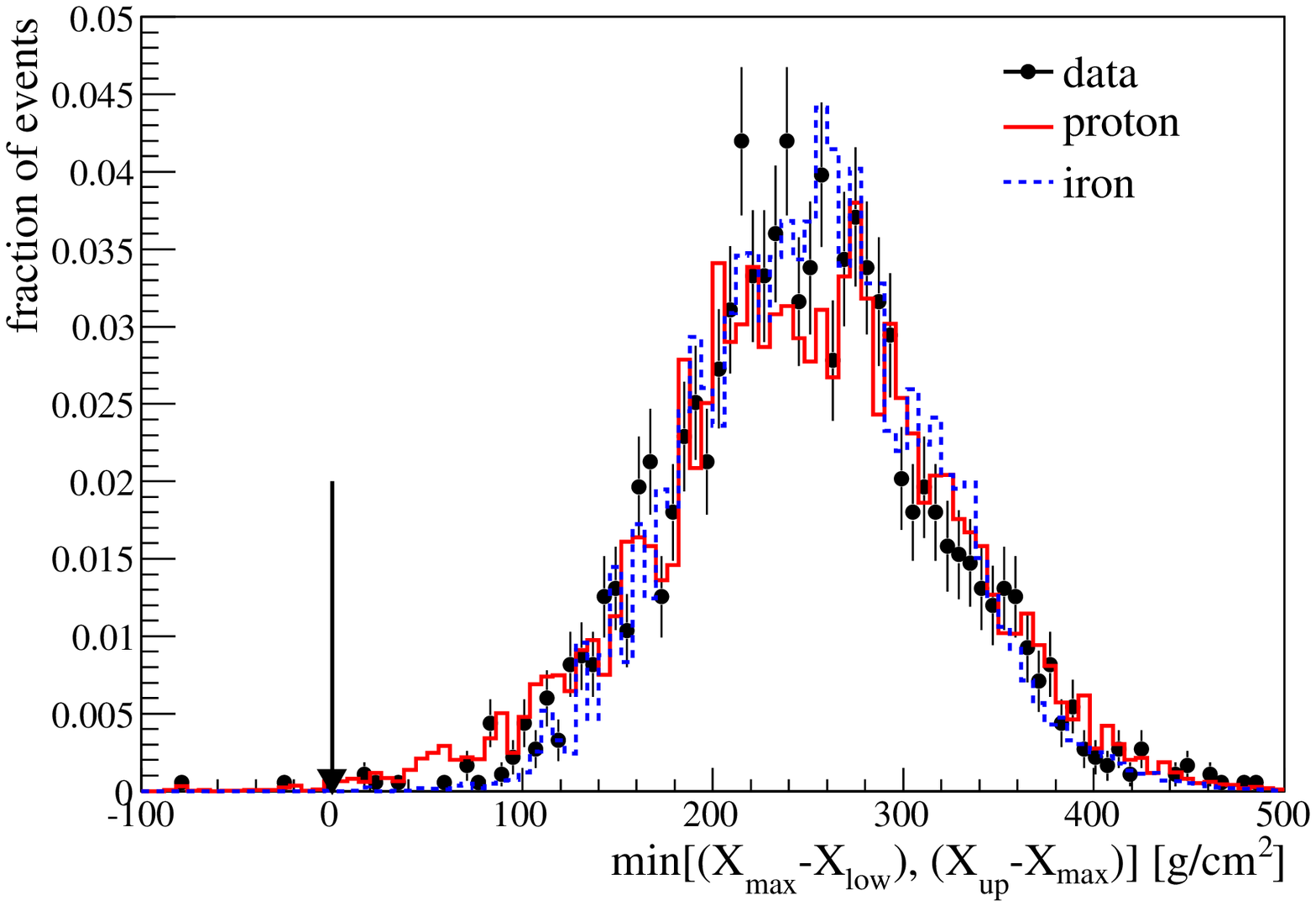}}
   \subfloat[Energy error]{\label{fig:EnergyError}
     \includegraphics[width=0.52\textwidth]{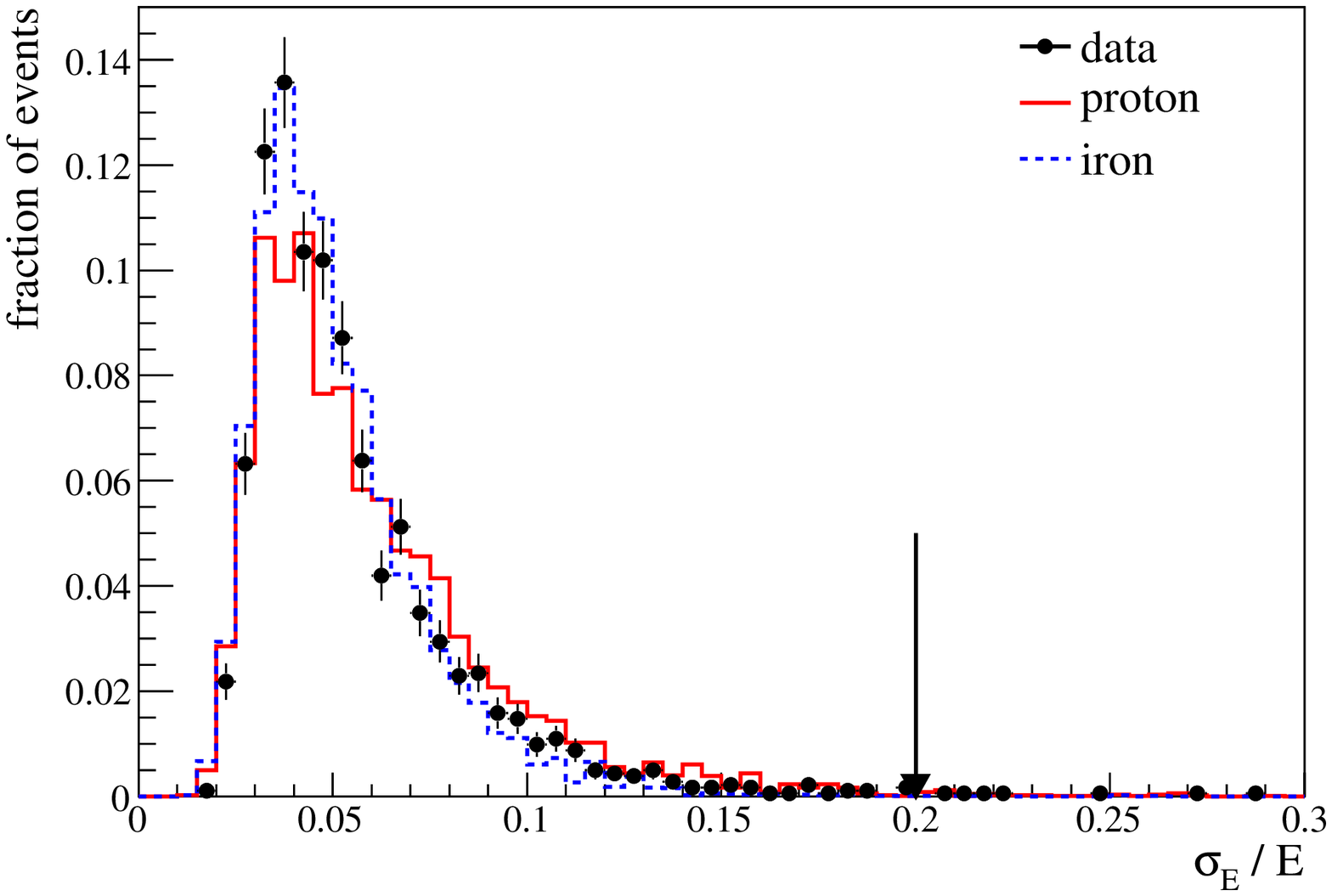}}
   \caption[]{Examples showing the agreement between simulation and
     data. Proton and iron primaries are shown separately for the
     simulated data. For each figure all quality and fiducial cuts
     are applied except the one related to the variable shown. The
     arrow denotes the selection cut on this variable.}
   \label{fig:MCvsData}
 \end{figure}

\subsection{Cross checks}
\label{sec:Crosschecks}

All the above criteria have been applied to both data and MC
events. The reliability of the quality criteria are checked by
comparing the cut parameter distributions of data and Monte
Carlo. Examples of these comparisons, shown in
figure~\ref{fig:MCvsData}, indicate a very good agreement between data
and Monte Carlo.

\begin{figure}[!t]
   \begin{center}
     \includegraphics[clip,width=0.95\textwidth]{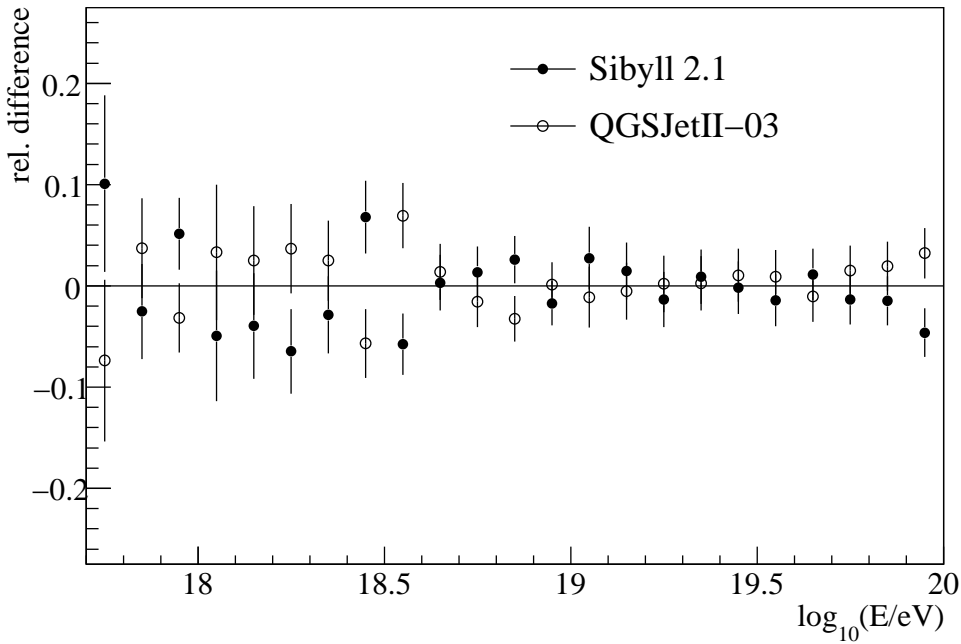}
        \caption{Relative difference between Sibyll and QGSJetII
          exposure with respect to the average of the two simulations. All
          quality and fiducial cuts are applied.}
        \label{fig:ModelDifference}
   \end{center}
\end{figure}

The exposure calculation depends somewhat on the hadronic interaction
model used in the Monte Carlo simulation.  Different hadronic
interaction models predict different fractions of shower energy
converted into visible light~\cite{Barbosa:2003dc} producing different
energy assignments and X$_{max}$ predictions. These differences might
affect the selection efficiency and lead to a model dependence in the
exposure. Two models, QGSJetII-03 and Sibyll 2.1 have been used as
input for the Time Dependent Detector Simulation and the selection
efficiencies have been compared. As is shown in
figure~\ref{fig:ModelDifference} the effect is lower than 2$\%$
averaged over the whole energy range. For this reason the exposure has
been calculated averaging the Monte Carlo samples simulated with the
different interaction models.

As mentioned above, the exposure is calculated as a function of
reconstructed energy to correct for distortions of the spectrum
introduced by the reconstruction of the shower energy.  It is well
known that this correction depends on the initial assumptions of the
true distribution (see eg.~\cite{Cowan}), i.e. on the energy
distribution of the generated events $N(E)$ in
Eq.~(\ref{eq.discrexposure}).  The exposure has been calculated using
different distributions for $N(E)$: power laws with three different
spectral indexes ($\gamma = 2,~3,~3.5$), a broken power law and the
combined spectrum measurement from the Pierre Auger
Observatory~\cite{AugerPLB2010}.  The ratio of the resulting exposure
$\mathcal{E}(E_\mathrm{rec})$ to the undistorted one,
$\mathcal{E}(E_\mathrm{gen})$, has been calculated for the different
cases. From this analysis, the choice of the input spectra used in the
Monte Carlo results in a systematic uncertainty lower than 2\%.

The availability in the Pierre Auger Observatory of two independent
detection techniques allows an overall validation of the Monte Carlo
simulation chain with data.  As shown in~\cite{SDTrAp} the surface
detector trigger is 100\% efficient above a few \EeV{}. Since SD data
are unaffected by the distance to the FD-site, light attenuation or
clouds, the FD trigger and selection efficiency can thus be measured
directly from the data.
A set of high quality SD showers have been selected during the time
periods with at least one FD-site taking data.  Given this set of
$N_\mathrm{SD}$ showers, we count the number of events that had at
least one triggered telescope, $N(\mathrm{FD_\mathrm{trig}})$, and
fulfilled all the selection criteria previously described,
$N(\mathrm{FD_\mathrm{sel}})$. The FD trigger and selection
efficiencies can then be estimated from:
\begin{equation} 
\varepsilon_\mathrm{trig} = P(\mathrm{FD_\mathrm{trig}}|\mathrm{SD}) =
\frac{N(\mathrm{FD_\mathrm{trig}})}{N_\mathrm{SD}} \end{equation} and \begin{equation}
\varepsilon_\mathrm{sel} = P(\mathrm{FD_\mathrm{sel}}|\mathrm{SD}) =
\frac{N(\mathrm{FD_\mathrm{sel}})}{N_\mathrm{SD}}.  
\end{equation}

\begin{figure}[!t]
   \begin{center}
     \includegraphics[clip,width=0.95\textwidth]{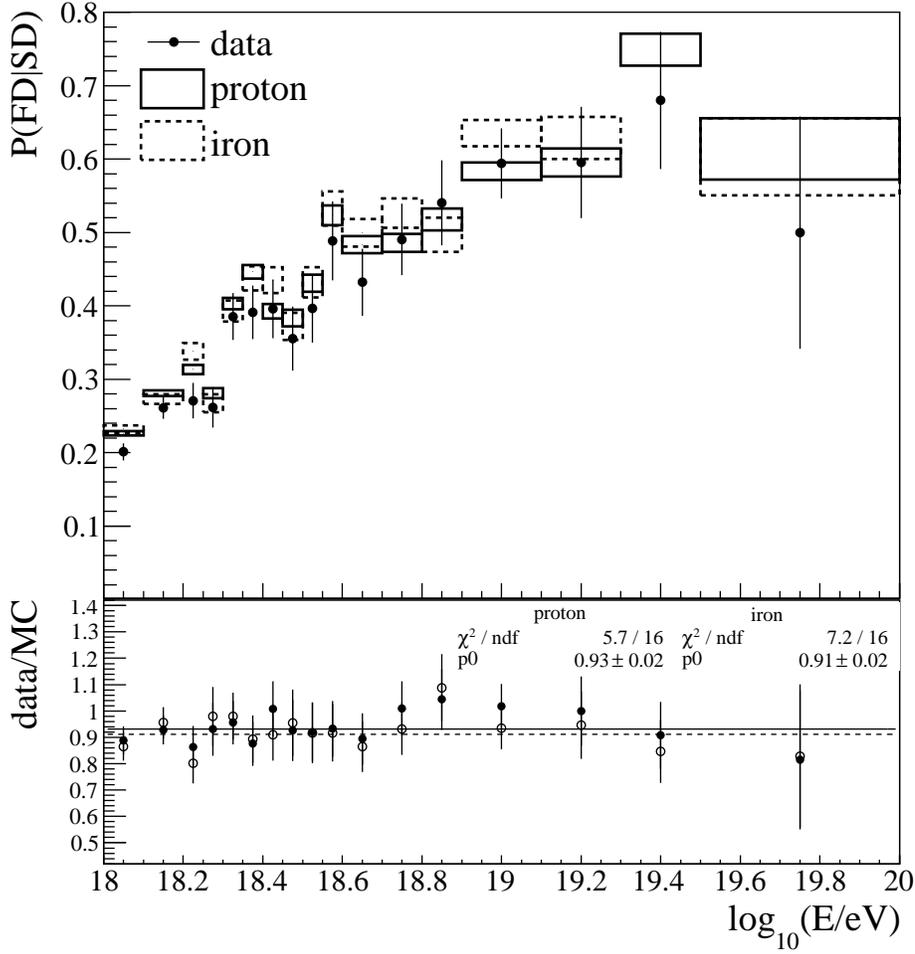}
        \caption{Conditional probability
          $P(\mathrm{FD}|\mathrm{SD})$. Comparison between measured
          and simulated efficiencies. The simulation has been
          performed using both proton and iron primaries.}
        \label{fig:fdsd}
   \end{center}
\end{figure}

For each data shower, 20 simulated CONEX showers are generated with
the given SD energy for proton and iron primaries. These showers are
then processed through the Time Dependent Detector Simulation with the
same arrival time, direction and core position as measured by the SD,
yielding the expected efficiencies:

\begin{equation}
\varepsilon^\mathrm{MC}_i = P(\mathrm{FD_i}|\mathrm{gen}) =
\frac{N(\mathrm{FD_i})}{N_\mathrm{gen}}, 
\end{equation} 

where $i$ stands for either the trigger or selection criteria and
$N_\mathrm{gen}$ denotes the number of generated events.

The Monte Carlo prediction is compared with the measurements in
figure~\ref{fig:fdsd}. It can be seen that the shape of the two curves
agree for both proton and iron simulations. However we note a
normalization factor between simulation and data of $0.92\pm0.02$
assuming a mixed composition of 50\% proton and 50\% iron nuclei. This
could be related to an uncertainty in the on-time, or caused by the
poorer energy resolution of the SD.

\section{Exposure}
\label{sec:exposure}
\begin{figure}[t]
  \begin{center} \includegraphics[width=0.95\textwidth]{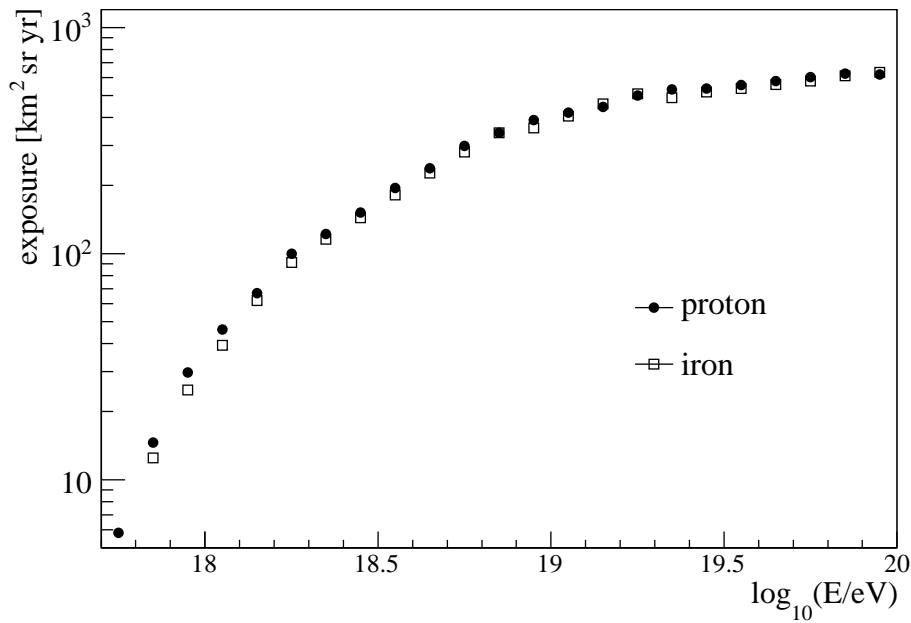} 
\caption{Hybrid exposure between November 2005 and May 2008 
for proton and iron primary particles.}
\label{fig:HybridExposure} 
\end{center}
\end{figure}

The hybrid exposure is shown in figure \ref{fig:HybridExposure} for
both proton (full circles) and iron (open squares) primaries. It is
calculated for the data period between November 2005 and May 2008, and
is that used for the hybrid energy spectrum measurement published
in~\cite{AugerPLB2010}.  The analysis of the Central Laser Facility
shots described in section \ref{sec:ontime} has revealed a systematic
shift in the on-time calculation. To take account of this effect the
exposure has been reduced by 4\%. Moreover the end-to-end comparison
in section \ref{sec:Crosschecks} has shown that the ratio of the true
event rate to that expected from Monte Carlo is $0.92 \pm 0.02$. The
systematic uncertainty of this comparison has been estimated to be
$\pm$ 5\%. Consequently the exposure has been reduced by half of the
corresponding correction ($\sim$~4\%) to cover the full range of
expectations. These two corrections are included in the exposure shown
in figure \ref{fig:HybridExposure}.

A mixed composition of $\unit[50]{\%}$ proton and $\unit[50]{\%}$ iron
nuclei has been assumed in the exposure
calculation~\cite{AugerPLB2010}. Numerical values of the hybrid
exposure can be found in \cite{hexptable}. The remaining composition
dependence has been included in the systematic uncertainty. This was
found to be about 8\% at 10$^{18}$ eV decreasing down to 1\% above
10$^{19}$ eV (figure~\ref{fig:primary_difference}b).  The dependence
of the exposure on the hadronic interaction model has been studied in
section \ref{sec:Crosschecks}. The effect is smaller than 2\% over the
entire energy range used for the calculation of the exposure. The
dependence of the exposure on the different input spectra used in the
Monte Carlo simulation has also been investigated and found to be
smaller than 2\%.  The overall systematic uncertainty on our knowledge
of the hybrid exposure has been obtained by summing all these
contributions in quadrature. It ranges from about 10\% at 10$^{18}$ eV
to 6\% above 10$^{19}$ eV.

\begin{figure}[t]
  \begin{center} \includegraphics[width=0.95\textwidth]{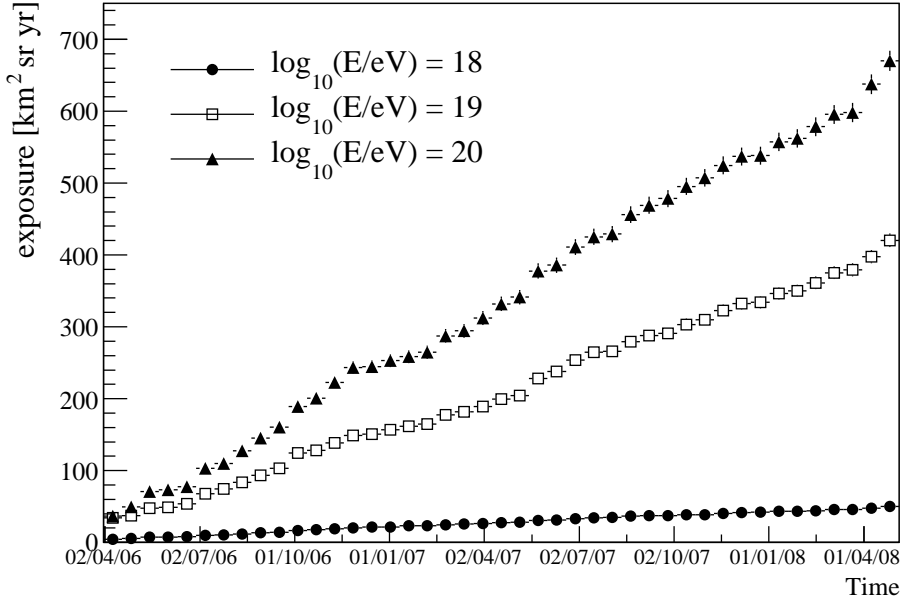} 
\caption{The growth of the hybrid exposure as a function of time starting from
    April 2006 up to May 2008 for three different energies.}
\label{fig:EvsT} 
\end{center}
\end{figure}

In figure \ref{fig:EvsT}, the growth of the hybrid exposure as a
function of time is shown for three different energies. The increase
with time shown at each energy comes as a result of the concurrence of
different effects, i.e the accumulation of data taking with time and
the growth of the SD array. One can also observe faster changes
corresponding to the longer FD data-taking periods in the austral
winter. The effect due to the growth of the SD array is more marked at
higher energies where a larger hybrid detection volume is accessible
with the new SD stations.

\section{Summary}
\label{sec:sum}
A method for the calculation of the hybrid exposure of the Pierre
Auger Observatory has been developed. The method is mainly based on
the Time Dependent Monte Carlo simulation technique. This technique
allows the simulation of a sample of events that reproduces in detail
the exact configuration of the experimental data taking, including
both instrumental and atmospheric conditions.

With this aim the on-time of the hybrid detector has been calculated
in a very accurate way taking into account environmental and
instrumental effects occurring at different levels of the DAQ process,
with respect to both the FD and the SD. The on-time information has
then been used as input for the Time Dependent Monte Carlo simulation.

In order to follow the fast changes of the detector configuration and
to get acceptable statistical and systematic uncertainties, a
simulation with very high statistics is crucial. A fast simulation has
been used, using CONEX shower profiles for the FD simulation and an
efficient simulation of the SD response. This fast approach has been
validated using a dedicated CORSIKA plus Geant4 simulation. No
significant difference has been found between the two approaches.

To obtain an unbiased measurement of the cosmic ray flux the exposure
estimate must be as free as possible of systematics. For this reason
only very high quality events have been used~\cite{AugerPLB2010}. To
satisfy this aim a set of quality criteria has been developed and
discussed in the paper. The effect of the quality criteria has been
cross-checked by comparing the data and Monte Carlo distributions; a
very good agreement has been found.

The systematic uncertainties arising from the unknown details of mass
composition, hadronic interaction physics and the true energy spectrum
have been calculated, and residual systematics from the on-time
calculation have been estimated. The overall systematic uncertainty on
the calculation of the hybrid exposure has been found to be lower than
10\% (6\%) at \energy{18} (above \energy{19}).

\section{Acknowledgements}
\label{sec:ack}
The successful installation and commissioning of the Pierre Auger Observatory
would not have been possible without the strong commitment and effort
from the technical and administrative staff in Malarg\"ue.

We are very grateful to the following agencies and organizations for financial support: 
Comisi\'on Nacional de Energ\'{\i}a At\'omica, 
Fundaci\'on Antorchas,
Gobierno De La Provincia de Mendoza, 
Municipalidad de Malarg\"ue,
NDM Holdings and Valle Las Le\~nas, in gratitude for their continuing
cooperation over land access, Argentina; 
the Australian Research Council;
Conselho Nacional de Desenvolvimento Cient\'{\i}fico e Tecnol\'ogico (CNPq),
Financiadora de Estudos e Projetos (FINEP),
Funda\c{c}\~ao de Amparo \`a Pesquisa do Estado de Rio de Janeiro (FAPERJ),
Funda\c{c}\~ao de Amparo \`a Pesquisa do Estado de S\~ao Paulo (FAPESP),
Minist\'erio de Ci\^{e}ncia e Tecnologia (MCT), Brazil;
AVCR AV0Z10100502 and AV0Z10100522,
GAAV KJB300100801 and KJB100100904,
MSMT-CR LA08016, LC527, 1M06002, and MSM0021620859, Czech Republic;
Centre de Calcul IN2P3/CNRS, 
Centre National de la Recherche Scientifique (CNRS),
Conseil R\'egional Ile-de-France,
D\'epartement  Physique Nucl\'eaire et Corpusculaire (PNC-IN2P3/CNRS),
D\'epartement Sciences de l'Univers (SDU-INSU/CNRS), France;
Bundesministerium f\"ur Bildung und Forschung (BMBF),
Deutsche Forschungsgemeinschaft (DFG),
Finanzministerium Baden-W\"urttemberg,
Helmholtz-Gemeinschaft Deutscher Forschungszentren (HGF),
Ministerium f\"ur Wissenschaft und Forschung, Nordrhein-Westfalen,
Ministerium f\"ur Wissenschaft, Forschung und Kunst, Baden-W\"urttemberg, Germany; 
Istituto Nazionale di Fisica Nucleare (INFN),
Istituto Nazionale di Astrofisica (INAF),
Ministero dell'Istruzione, dell'Universit\`a e della Ricerca (MIUR), Italy;
Consejo Nacional de Ciencia y Tecnolog\'{\i}a (CONACYT), Mexico;
Ministerie van Onderwijs, Cultuur en Wetenschap,
Nederlandse Organisatie voor Wetenschappelijk Onderzoek (NWO),
Stichting voor Fundamenteel Onderzoek der Materie (FOM), Netherlands;
Ministry of Science and Higher Education,
Grant Nos. 1 P03 D 014 30 and N N202 207238, Poland;
Funda\c{c}\~ao para a Ci\^{e}ncia e a Tecnologia, Portugal;
Ministry for Higher Education, Science, and Technology,
Slovenian Research Agency, Slovenia;
Comunidad de Madrid, 
Consejer\'{\i}a de Educaci\'on de la Comunidad de Castilla La Mancha, 
FEDER funds, 
Ministerio de Ciencia e Innovaci\'on and Consolider-Ingenio 2010 (CPAN),
Generalitat Valenciana, 
Junta de Andaluc\'{\i}a, 
Xunta de Galicia, Spain;
Science and Technology Facilities Council, United Kingdom;
Department of Energy, Contract Nos. DE-AC02-07CH11359, DE-FR02-04ER41300,
National Science Foundation, Grant No. 0450696,
The Grainger Foundation USA; 
ALFA-EC / HELEN,
European Union 6th Framework Program,
Grant No. MEIF-CT-2005-025057, 
European Union 7th Framework Program, Grant No. PIEF-GA-2008-220240,
and UNESCO.

\bibliographystyle{ieeetr}
\bibliography{HybridExposure}

\end{document}